# Treatment of Semantic Heterogeneity in Information Retrieval

Heiko Hellweg, Jürgen Krause, Thomas Mandl, Jutta Marx, Matthias N.O. Müller, Peter Mutschke, Robert Strötgen

InformationsZentrum
Sozialwissenschaften
(Social Science Information Centre)

Lennéstraße 30
D-53113 Bonn
Tel.: 0049 228 22810
Fax.: 0049 228 2281120
email: marx@bonn.iz-soz.de
Internet: http://www.gesis.org

# Content

# 1 Introduction

## 1.1 What Are Transfer Relations and How to Use Them

Nowadays, users of information services are faced with highly decentralised, heterogeneous document sources with different content analysis. Semantic heterogeneity occurs e.g. when resources using different systems for content description are searched using a single query system. This semantic heterogeneity is much harder to deal with than the technological one. Standardization efforts such as the Dublin Core Metadata Initiative (DC) are a useful precondition for comprehensive search processes, but they assume a hierarchical model of cooperation, accepted by all players.

Because of the diverse interests of those partners, such a strict model can hardly be realised. Projects should consider even stronger differences in document creation, indexing and distribution with increasing „anarchic tendencies"(cf. Krause 1996, Krause/Marx 2000). To solve this problem, or at least to moderate it, we suggest a system consisting of an automatic meta-data generator and a couple of transformation modules between different document description languages. These special agents need to map between different thesauri and classifications.

The first step to handle semantic heterogeneity should be the attempt to enrich the semantic information about documents, i.e. to fill up the gaps in the documents meta-data automatically. Section 2 describes a set of cascading deductive and heuristic extraction rules, which were developed in the project CARMEN for the domain of Social Sciences.

The mapping between different terminologies can be done by using intellectual, statistical and/or neural network transfer modules. Intellectual transfers use cross-concordances between different classification schemes or thesauri. Section 3 describes the creation, storage and handling of such transfers.

Statistical transfer modules can be used to supplement or replace cross-concordances. They allow a statistical crosswalk between two different thesauri or even between a thesaurus and the terms of automatically indexed documents. The algorithm is based on the analysis of co-occurrence of terms within two sets of comparable documents. The main principles of this approach are discussed in section 4.

A fundamental problem of co-occurrence analysis is to find documents of similar content, i.e. to build up a parallel corpus. Because this cannot be done in all domains, a corpus has to be simulated in some. This simulation is explained in section 5.

Section 6 describes the handling of semantic heterogeneity by using neural networks, especially the backpropagation model.

The traditional form of vagueness treatment in Information Retrieval refers to the comparison between query terms and content analysis terms, whereby the document level is regarded as the uniform modelling basis. Opposed to this, the heterogeneity modules mentioned above are used within the so called "two-step" model (cf. Krause 2000), which was developed at the Social Science Information Centre (IZ) in the context of the projects ELVIRA, CARMEN ViBSoz, and ETB (cf. section 1.2).

It is based on the assumption, that heterogeneous document sets should first be interlinked through transfer modules (vagueness modelling at the document level) before they are integrated in the super ordinate process of vagueness treatment between documents and query (the traditional Information Retrieval problem).

If, for example, three heterogeneous document sets have to be integrated, transfer modules between A - B and B - C, each bilaterally treat the vagueness between the different content analysis methods (cf. Figure 1). The hope behind this form of vagueness treatment, which differs considerably from the procedure traditionally used in Information Retrieval, is to produce greater flexibility and target accuracy of the overall procedure through separation of the vagueness problem. Different forms of vagueness do not flow into one another uncontrolled, but can be treated close to their cause (e.g. the differences between two thesauri). This appears more plausible in cognitive terms and permits the combination of a wide range of modules for treatment of vagueness. This combination of modules becomes quite effective, when applied on the retrieval of heterogeneous data sets.

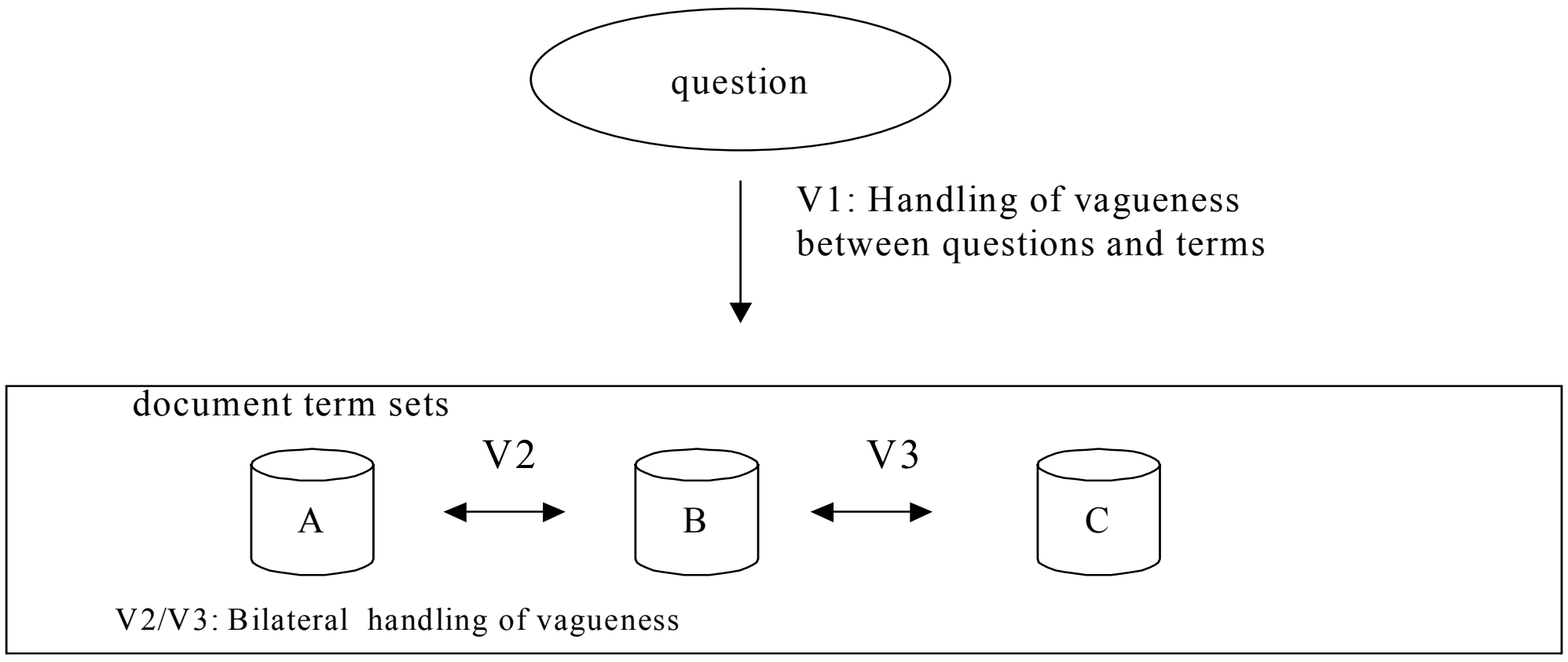

**Figure 1: Two-step method**

Section 7 describes the technical realisation of the two-step method in concrete information retrieval systems.

## 1.2 IZ Projects and Their Domains

### 1.2.1 ELVIRA

The project ELVIRA (ELektronisches Verbands InfoRmAtionssystem, electronic industry association information system) was funded by the German Ministry of Commerce and Technology (BMWI, grant# VI B4 003060/22) and industry partners, namely the industry associations ZVEI (an association of the electronic manufacturers industry), VDMA (an association of the machine manufacturers industry) and HVB (an association of the building industry). During the first funding period (from January 1995 to December 1996) a dedicated retrieval system of multi-categorized factual data was designed and implemented. In the second period (April 1997 to June 2000) it was augmented with a retrieval component for textual data from different sources, containing heterogeneous sets of meta-information for the different text bases. Both were tightly coupled to allow the search for data of all types using automatic query transformation.

Factual data like timelines are categorized according to three facets (e.g. topic, business and region). A dynamic user interface allows for the simplified navigation and retrieval of the actual data by employing knowledge about the available data pools and their relations.

Textual data with different flavors of metadata from different sources (in house texts from the industry associations, as well as information from commercial agencies) are organized in data pools. A text retrieval component allows the formulation of queries for different types of data pools in a dynamically integrated user interface. Form based queries and logical (i.e. Boolean) expressions are integrated with advanced search techniques like probabilistic full text search with ranking.

Query transformation was implemented to support not only the simplified search on multiple sets of text data but also to permit the integration of textual and factual data. Queries for fact data can be transformed into queries for text data and vice versa. This allows for a previously unknown tight coupling of data sources with very different structures.

### 1.2.2 ViBSoz

The Social Science Virtual Library project[1] (funded by the 'Deutsche Forschungs Gemeinschaft, DFG') started in May 1999 and will end in May 2001. It aims at presenting an integrated view to distributed, heterogeneous data of German social science literature. There are some special libraries, some general libraries with large amounts of social science literature and one central bibliographic database (SOLIS). All of these not only have different user interfaces but worse, they use different systems of content description like thesauri and classifications. The goal of the project is to give the user a central access point with a single user interface and an integrated view to thesauri/classifications. The main aspect is to give the user the ability to search different databases which have different vocabulary for content analysis with just one query and to present the overall result in a single, uniform result set.

### 1.2.3 CARMEN

The project CARMEN ("Content Analysis, Retrieval and MetaData: Effective Networking")[2] is funded by the German Ministry for Education and Research in the context of the "Global Info" program. It started on 1 October 1999 and will end on 28 February 2002.

1 http://www.gesis.org/Forschung/Informationstechnologie/ViBSoz.htm
2 http://www.mathematik.uni-osnabrueck.de/projects/carmen/

The project aims at finding a suitable information system for the distributed data collections in libraries, information centres and the Internet for the domains mathematics, physics and social sciences. The idea behind this effort is to combine different information sources to solve the useful information-finding problem in the “borderless world of the Internet”. In the project’s context this task is more a semantic and conceptual problem than a technical one. Heterogeneity appears in different data collections using different thesauri or classifications for content description, when they contain varying or no metadata at all, or when intellectually indexed documents meet completely un-indexed Internet pages. Standardization, such as the Dublin Core (DC) metadata elements, is a useful precondition for comprehensive search processes but it assumes a hierarchical model of cooperation, accepted by all players. This strict model can hardly be realized in short term. Therefore we work on a system of automatic transfer modules, which map the user’s query to the underlying metadata of each database in question syntactically and semantically as an additional approach to improve the retrieval. The specific task of this project compared with ELVIRA and ViBSoz is combining not only library catalogues and specialised information sources, but the wide range of un-indexed Internet documents, too.

Three working groups “Metadata”, “Retrieval” and “Heterogeneity” handle main aspects of this task. The University of Osnabrück and the German Social Science Information Centre work cooperatively on the task “Treatment of heterogeneity for textual information of different data types and content analysis tools”. This work package creates statistical transfer relations and extractors for metadata extraction from Internet documents.[3]

The project partners for the work package “Concordances of classifications and thesauri” are “Die Deutsche Bibliothek”, Frankfurt, the University Library of Regensburg, the German Institute for International Educational Research, Frankfurt, the Max Planck Institute for Human Development, Berlin, and the IZ. This work package produces intellectual transfer relations.

### 1.2.4 ETB

The project European Schools Treasury Browser (ETB) is founded by the European Commission and embedded into the European Schoolnet (EUN)[4] where nearly all European countries are members through their ministries of education (not only member states of the EU, but also those applying for membership, being associated or somehow linked to the EU). The ETB project[5] is running from February 2000 to March 2002. It aims at the establishment of a network of networks of educational repositories in Europe and at the offer of an educational portal for all school related online resources[6]. To achieve this goal the problems of multilinguality and heterogeneity have to be overcome. Besides the development of a multilingual thesaurus methods and instruments to handle the different layers of content description shall be developed.

---

3 http://www.bonn.iz-soz.de/research/information/carmen/ap11/

4 http://www.eun.org

The data which are held by the repositories which will be included in the network vary with respect to the language, the metadata obtained, and the content description methods used (indexing by thesaurus terms, free indexing, use of classification schemes, no content description at all). The repositories contain information on teaching materials, school projects, class room projects, and they are reaching from textual information to multimedia and interactive programs.

The ETB project plan asks for building cross-concordances between relevant classification schemes and between relevant thesauri, and for developing transfer modules for augmenting the content descriptions. The use of mapping classification schemes and thesauri by developing intellectually made cross-concordances and statistically based transfer modules is supposed to solve the problems of semantic heterogeneity and to assist the multilingual access.

For solving the problems of multilinguality the same methods as for the solution of general problems of semantic heterogeneity can be used. The work on cross-concordances can not only be applied to different thesauri in one language but also to thesauri in various languages. The work on transfer modules can also be applied to the relationships between languages.

Inside the ETB project not all strategies which have been used for the other contexts of the research of the IZ (CARMEN, ViBSoz, ELVIRA) are replicated: neural networks and metadata extraction will not be used in the ETB context because they are much more domain dependent and need a lot of more training data, comparable or parallel corpora, and test runs for establishing reliable solutions. These methods would have needed much more in depth work which is not possible within the restricted resources and time of the project.

---

5 http://etb.eun.org

6 Overview of the project by Kluck (2001)

The specific problems of using cross-concordances and transfer modules in the ETB project is the big heterogeneity of repositories and their structures. Some, but few repositories offer a description of the selected links which is as well available as plain text as well augmented with metadata tags in the source text. Some repositories offer their content description in a database which does not make the metadata labels accessible, they have to be extracted from the database, but cannot be checked online. Most repositories offer only links (partly with short annotations from the source documents) to interesting resources.

# 2 Extraction of Meta-Data from Internet Documents

The goal of extracting meta-data during the gathering process[7] is to enrich poorly indexed documents with additional meta-information, e.g. author, title, keywords or abstract. This meta-data is meant to be used in retrieval in the same way as intellectually added meta-data yet with lower relevance.

The actual algorithms for meta-data extraction depend on file formats, domain properties, site properties and style properties (layout). No stable and domain independent approach is known so far. Until conventions for creating HTML documents change and become standardized towards a semantic web only temporary and limited solutions can be found.

In the context of the project CARMEN for example (cf. Strötgen/Kokkelink 2001), relevant documents from the mathematics domain exist mostly in PostScript format. These unstructured documents (thesis papers) contain some meta-data of high value, which are marked only by format information (e.g. font size) or special keywords. By using and modifying some tools like prescript from the New Zealand Digital Library[8], a sample of PostScript documents has been transformed and analysed. Abstract, keywords and classification terms can be extracted from these documents successfully; identification of author and title information is still work-in-progress.

---

[7] While the other methods described in the following sections (cf. sections 3-6) run during the indexing or retrieval process, this method is integrated into the gathering process. The extracted meta-data is attached to the document and indexed with additional weight information like other meta-data.

[8] http://www.nzdl.org/

Internet documents from the social sciences are mostly structured HTML files, but they use HTML features mainly for formatting, not as mark-up tags for content. Meta tags can rarely be found, and even more or less correct HTML syntax cannot be relied upon. Different institutions use very different ways of creating their Internet documents and a large number of documents contain no information about author or origin at all. This makes extraction of meta-data very difficult.

Nevertheless we developed heuristics for identifying some meta-data in this heterogeneous set of documents. Because operating on (frequently incorrect) HTML files is not very comfortable, we transformed the documents into (corrected) XHTML and implemented our heuristics on these XML trees using XPath queries.

The following algorithm (<x> shows an internal method number, [x] the weight of the extracted meta-data) processes the parts of the document that have been identified and extracted with XPath before (cf. Figure 2).

This algorithm generates a useful and weighted title for most Internet documents found for social sciences. We do not create specialized extractors for single sources or wrappers because we do not have enough resources to create and maintain them. However, we have general extractors that generate reasonably applicable meta-data, which is of higher quality than the existing poor indexed documents.

First tests show that the extractor for plain documents from the domain mathematics produces very good results for abstracts (nearly 90%) as well as for keywords and classifications (more than 90%). The extractions for the domain social sciences generate very good results for the title (about 90%) but are less successful for keywords and abstracts (less than 15%). And of course the extracted meta-data cannot be more reliable than the original document – a misleading title given by the author remains worthless after the extraction process.

```
Extraction of relevant data with XPath-queries:

1. Find first <title> tag in the document (XPath: //title)
2. Let H1=(x_1,..,x_n) be the node list of query //h1.
   Combine the sequence x_l,…,x_k with (x_i is the previous neighbour of
   x_i+1) to one single item. (Thus H1 is a list of items with no x_i being
   direct previous neighbour of x_i+1.)
   Proceed this for H2,…,H6 analogously. The result is a matrix H =
   (1, string1)
   (1, string2)
   (2, string3) and so on. At this the first item is the grade of the head-
   ing. Let H1 be the first item (1,H1) in H; analogous for H2,…,H6. HMAX
   is the heading with the highest grade.
3. Search for all combinations //p/b , //b/p etc. (special list of empha-
   sized paragraphs).
   Let s_1,…,s_n be a nodelist made up of the queries:
     //*[self::p/parent::strong or self::p/parent::b or
     […]
     self::i/parent::td]
   This list is ordered by the appearance in the document.
   Find the highest k with for any s_I from {s_1,…,s_k} applies:
   1. name(s_i)=name(s_(i+1))
   2. name(parent(s_i))=name(parent(s_i+1))
   3. parent(s_i) is the previous neighbour of parent(s_(i+1)).
   Combine this s_1,…,s_k to one String S.

Heuristic:

If (<title>-tag exists && <title> does not contain "untitled" && HMAX
    exists){
    /* 'does not contain "untitled"' is to be searched as case
       insensitive substring in <title> */
    If (<title>==HMAX) {
      <1> Title[1]=<title>
    } elsif (<title> contains HMAX) {
      /* ' contain' does always mean case insensitive substring */
      <2> Title[0,8]=<title>
    } elsif (HMAX contains <title>) {
      <3> Title[0,8]=HMAX
    } else {
      <4> Title[0,8]=<title> + HMAX
    }
  } elsif (<title> exists && S exists) {
    /* i.e. <title> exists AND an item //p/b, //i/p etc. exists */
    <5> Title[0,5]=<title> + S
  } elsif (<title> exits) {
    <6> Title[0,5]=<title>
  } elsif (<Hx> exits) {
    <7> Title[0,3]=HMAX
  } elsif (S exits) {
    <8> Title[0,1]= S
  }
}
```

**Figure 2: Heuristic for Title Extraction.**

# 3 Intellectual Transfer Relations

Intellectual transfers use cross-concordances between different controlled documentation languages like thesauri and classifications. These languages have a limited number of indexing terms or classes used to describe document subjects. A thesaurus is a natural language based documentation language with different kinds of vocabulary control and terminological control (i.e. synonym control and homonym control). Every thesaurus term is unique. Relations like equivalence, hierarchical and associative relations are defined between the descriptors (cf. Burkart 1997). A classification is usually an artificial (alphabetical, numeric or alphanumerical) documentation language. A classification has classes or notions, which are systematically ordered with hierarchical relations, the classification structure. The class and its concept have a verbal class description (cf. Manecke 1997).

Documentary and professional experts create semantic relations between thesaurus terms or classes with similar meanings.

Different kinds of inter-thesaurus or classification relations are defined: “exact equivalence”, “inexact equivalence”, “narrower equivalence” and “broader equivalence”. The relations have weight information (“high”, “medium”, “low”) reflecting the relevance of the relations for retrieval quality.

In the project CARMEN, for example, cross-concordances are provided between universal or special classifications and thesauri in the domains involved (mathematics, physics, social sciences). These cross-concordances allow safe transfers between different documentation languages and the document sets using them.

Problems may arise if there are insufficient resources (time, money, domain experts) to create such cross-concordances; furthermore not all documents – particularly Internet documents – are indexed with a controlled vocabulary. Therefore additional approaches like statistical transfers based on the analysis of co-occurrence of terms (cf. sec. 4) are necessary.

The software tool SIS-TMS[9] proved useable for creation of cross-concordances between different thesauri. CarmenX[10] has proved to be equally useful for creating relations between different classifications.

---

[9] http://www.ics.forth.gr/proj/isst/Systems/sis-tms.html

## 3.1 SIS-TMS

SIS-TMS is a system to create and maintain multilingual thesauri and cross-concordances produced by the Institute of Computer Science, Foundation for Research and Technology - Hellas (ICS-FORTH). It can be used to modify or create semantic relation types within and between thesauri, and for presentation of views of the network with coloured graphs. It is therefore a tool to analyse the logical structure and consistency of thesauri and a solution to manage a set of correlated terminology resources for classification and data access in a distributed environment.

SIS-TMS consists of a terminology server for integration with heterogeneous electronic collections, a graphical browser for cataloguers and end-users (cf. Figure 3), and a tool to cooperatively develop multilingual thesauri. Some main features of the TMS are development and access to multiple thesauri and their interrelations; creating views thereon; managing partial releases; and creating consistent update increments for federated collection databases.

The SIS-TMS graphical user interface allows the navigation within and between different thesauri and the execution of predefined queries and the presentation of graphical views to identify concepts for cataloguing or database queries, to identify translations or equivalent expressions for information access in heterogeneous environments, and to control the quality and logical consistency of a system of interlinked thesauri during development. Configuration tables allow for localization.

The SIS-TMS server can be integrated into a distributed, heterogeneous environment. It allows access through its programmatic interface. It further permits automatic term expansion and translation into queries of distributed data sources.

The SIS-TMS system is an application of the Semantic Index System[11], a general-purpose object-oriented semantic network database. Its schema includes the principles of the ISO 2788 and 5964 standards for monolingual and multi-lingual thesauri. The schema can be changed and extended by user definition in order to adapt SIS-TMS to the special documentation languages and user needs. The schema for thesauri was expanded to also handle classifications. Classifications in the enhanced SIS-TMS schema have no internal relations, other than hierarchical ones. The descriptive text for a class respectively notion is defined as a translation to a String in German or another language. SIS-TMS can read the classifications and classification cross-concordances and gives the transfer modules a common path for access to the database.

---

10 http://www.bibliothek.uni-regensburg.de/projects/carmen12/index.html.en
11 http://www.ics.forth.gr/proj/isst/Systems/sis.html

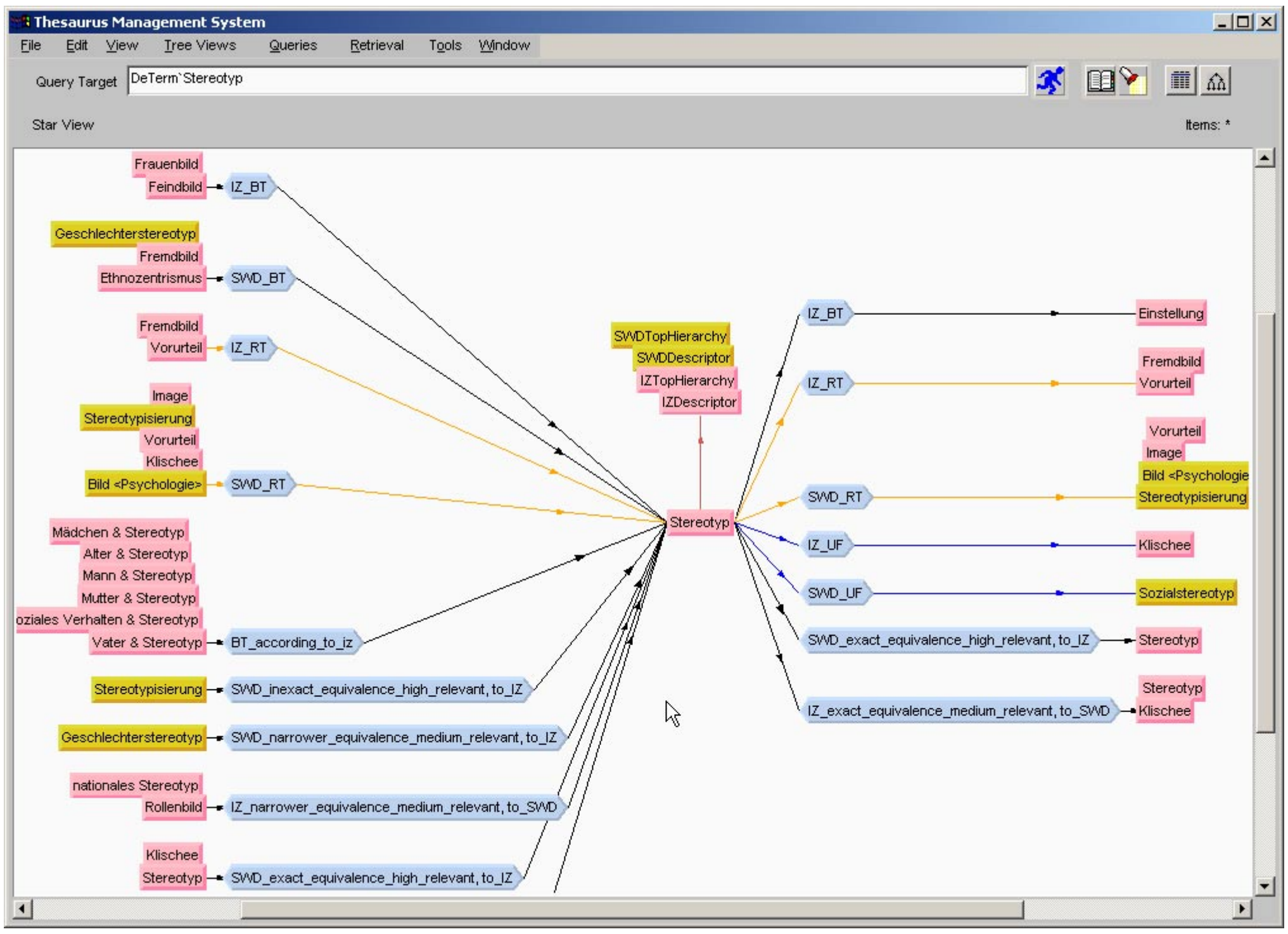

**Figure 3: „Star View" with all inner and inter thesaurus relations for the term "Stereotyp"**

A SIS-TMS server runs in the IZ and is free for access by other CARMEN project partners. The online access to SIS-TMS requires a fast network connection; access within the German scientific broadband network (G-WiN) allows tolerable response times.

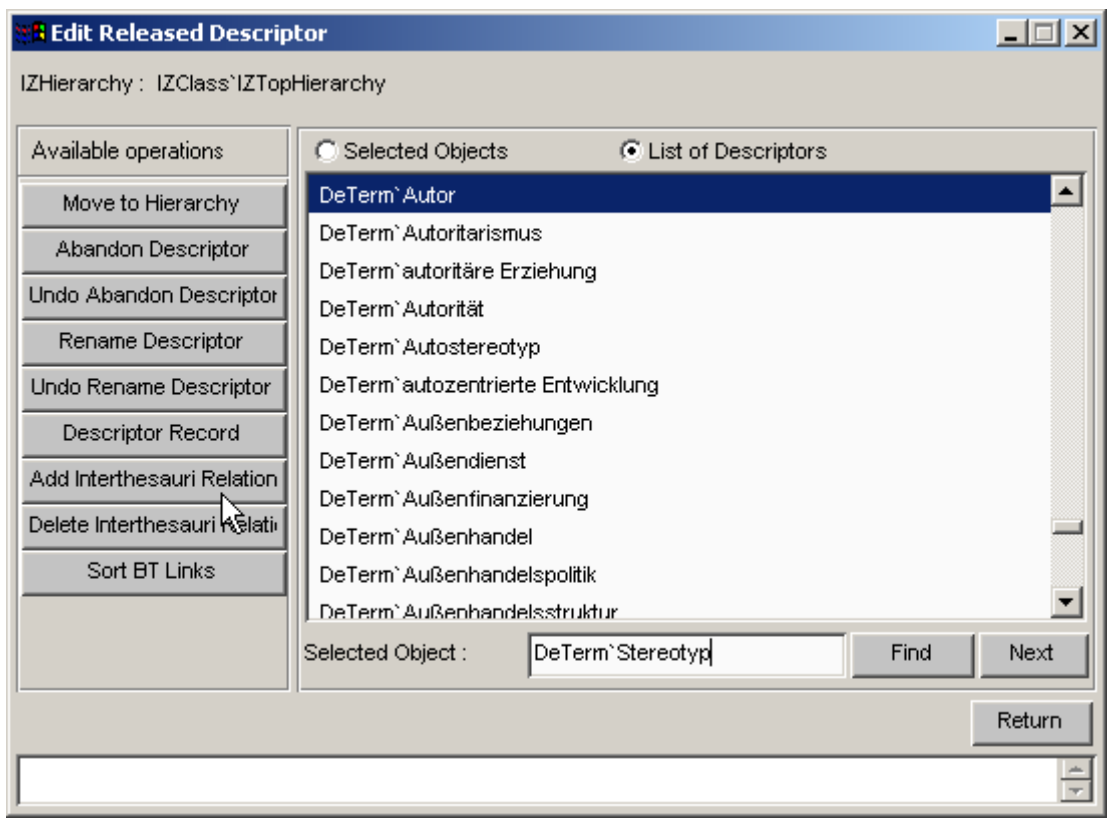

**Figure 4: Entry Form for creating and editing thesaurus terms and relations**

The “Entry Form” for creating and manipulating thesauri terms and relations (cf. Figure 4) proved to be clumsy to use (cf. Figure 5) and to be faulty in some aspects. There is no consistent integration between the thesaurus browser and the entry tool that allows a coherent workflow for data manipulation. Therefore the module is not currently in use; relation tables created off-line are parsed and converted to TELOS descriptions (the SIS language for manipulations of schemes and elements in the semantic network). We read these TELOS descriptions into the SIS database and allow online access to it for browsing and for querying the database. Improvements and corrections of the “Entry Forms” will hopefully allow online manipulation of the network very soon.

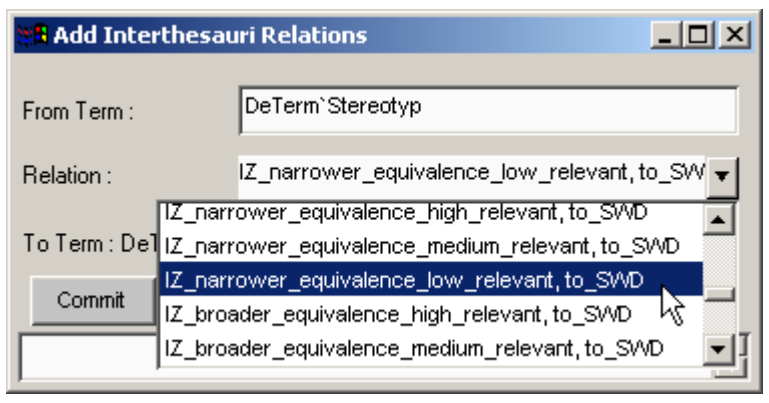

**Figure 5: Entry Form for creating inter thesaurus relations**

The graphical tool for browsing thesauri and relations offers some relevant views on the database (cf. Figure 6). It is fairly customisable, but it has some limitations and bugs, which will be probably removed in near future. This tool is a stand-alone application that regrettably can not be integrated into a custom user interface for supporting query strategies or clarifying retrieval results.

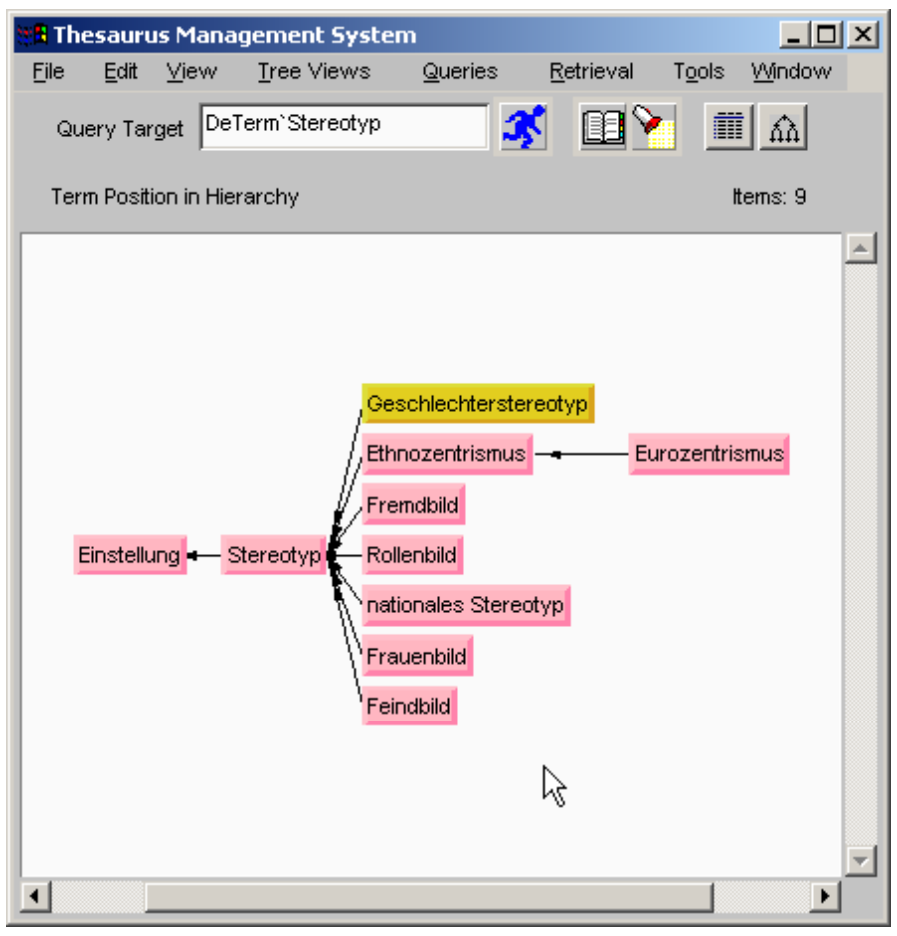

**Figure 6: „Term Position in Hierarchy View“ with hierarchical inner thesaurus relations for the term “Stereotyp”**

## 3.2 CarmenX

CarmenX is a tool for creating cross-concordance relations for classifications developed by the University Library of Regensburg especially for the CARMEN project. It is World Wide Web based and can be used with a 4th generation browser like Netscape Communicator 4.x or Microsoft Internet Explorer 4.x/5.x as client (cf. Figure 7). The server is implemented by a http server with Apache, Perl and PHP; the concordances are stored in a MySQL database. The classifications themselves can be stored in different locations spread over the Internet. Cooperative and distributed work on the cross-concordances is possible.

CarmenX opens two classifications in two side by side tree views. Browsing the corresponding classes in each tree and producing the relation with type and weight information can create relations between classes. The links are created in one direction only and can therefore be defined asymmetrically, but the opposite direction of the relationship can be added by running a batch program. CarmenX displays and allows editing of existing links.

A short description of the CarmenX tool is available in German and English[12].

CarmenX is designed for creating cross-concordances for classifications only and not for thesauri. It may be extended for use with thesauri in near future.

[12] see at http://www.educat.hu-berlin.de/~kluck/carmenx-etb-de.rtf (German) and at http://www.educat.hu-berlin.de/~kluck/carmenx-etb-en.rtf (English)

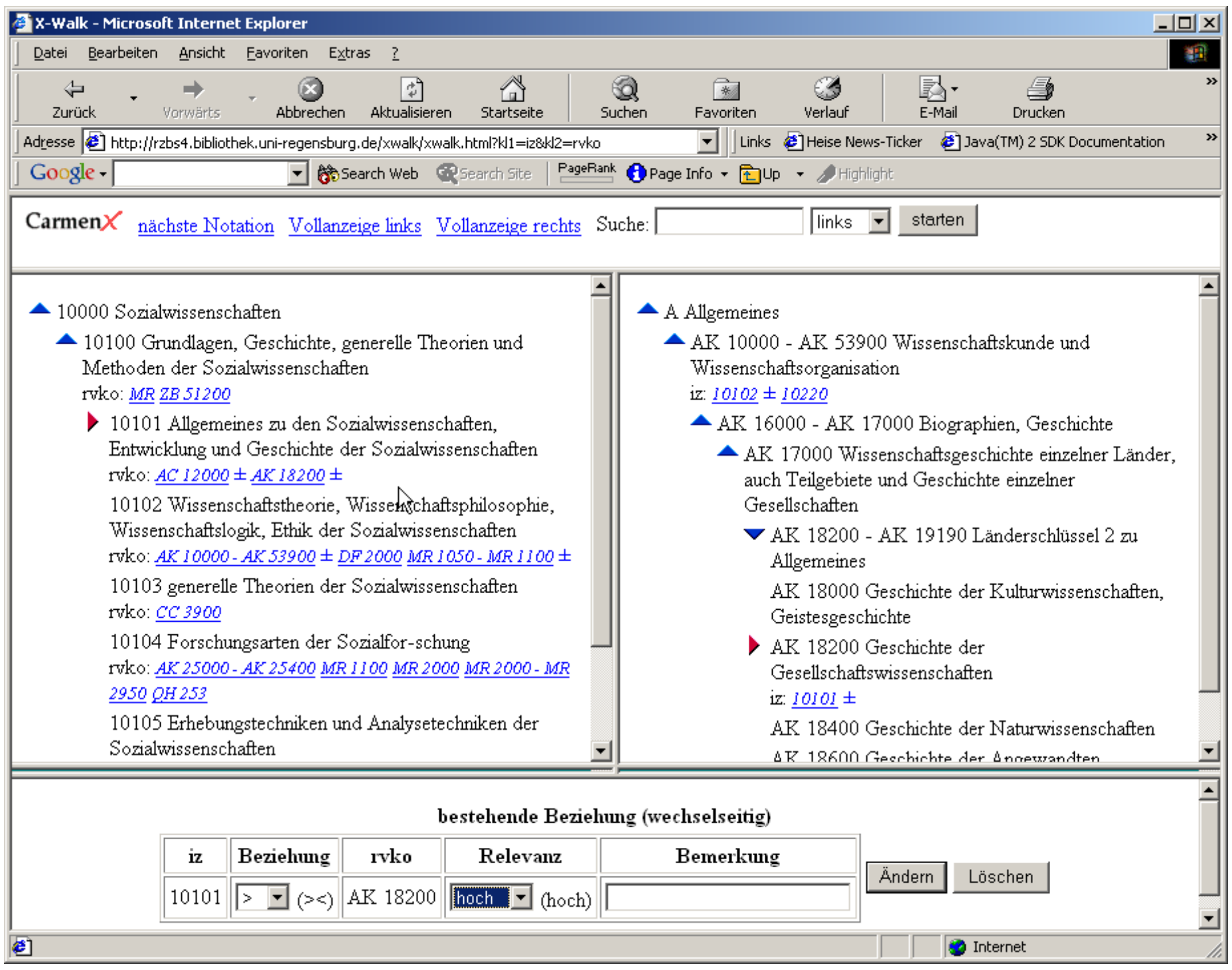

**Figure 7: Classification trees for IZ Classifications and RVK in CarmenX**

# 4 General Principle of Co-Occurrence for Statistical Transfer Relations

Intellectual cross-concordances (as described in section 3) are very precise transfer relations. They allow different types of relations to be combined with a high quality overall result. Their major disadvantage is their cost. As the result of an intellectual process they require considerable human resources to create, and as such they are simply expensive. Not all institutions can afford them, or are willing to commit the necessary resources. Therefore other – cheaper – ways had to be found to create transfer relations.

Quantitative statistical methods offer a general, automatic way to create transfer relations on the basis of actual bibliographic data. Co-occurrence analysis is one of those methods. It takes advantage of the fact that the content analysis from two different libraries a single document hold in both collections will represent the same semantic content in different content analysis systems. The terms from content analysis system A that occur together with terms from content analysis system B can be computed. The assumption is that the terms from A have (nearly) the same semantics as the related ones from B, and thus the relation can be used as a transfer relation. The prerequisite for such computations is a parallel corpus, where each document is indexed by two different content analysis systems.

## 4.1 Parallel Corpora

The classical parallel corpus is based on two different sets of documents, e.g. two different library catalogues. Each is indexed with a specific thesaurus/classification. To be able to create co-occurrence relations between the terms of these thesauri, the indexations of the documents have to be brought into relation. This is done by finding identical (or at least equivalent) documents in both catalogs. Considering print media, the problem of identity can be solved quite easy. An identical ISBN in combination with at least one identical Author should be a sufficient criterion for the identity of two documents. But the situation is much less easy, if the underlying data sets are not bibliographic ones, e.g. if text data should be combined with fact data or if internet texts are considered. As we will see later, other ways of creating a parallel corpus have been suggested for this.

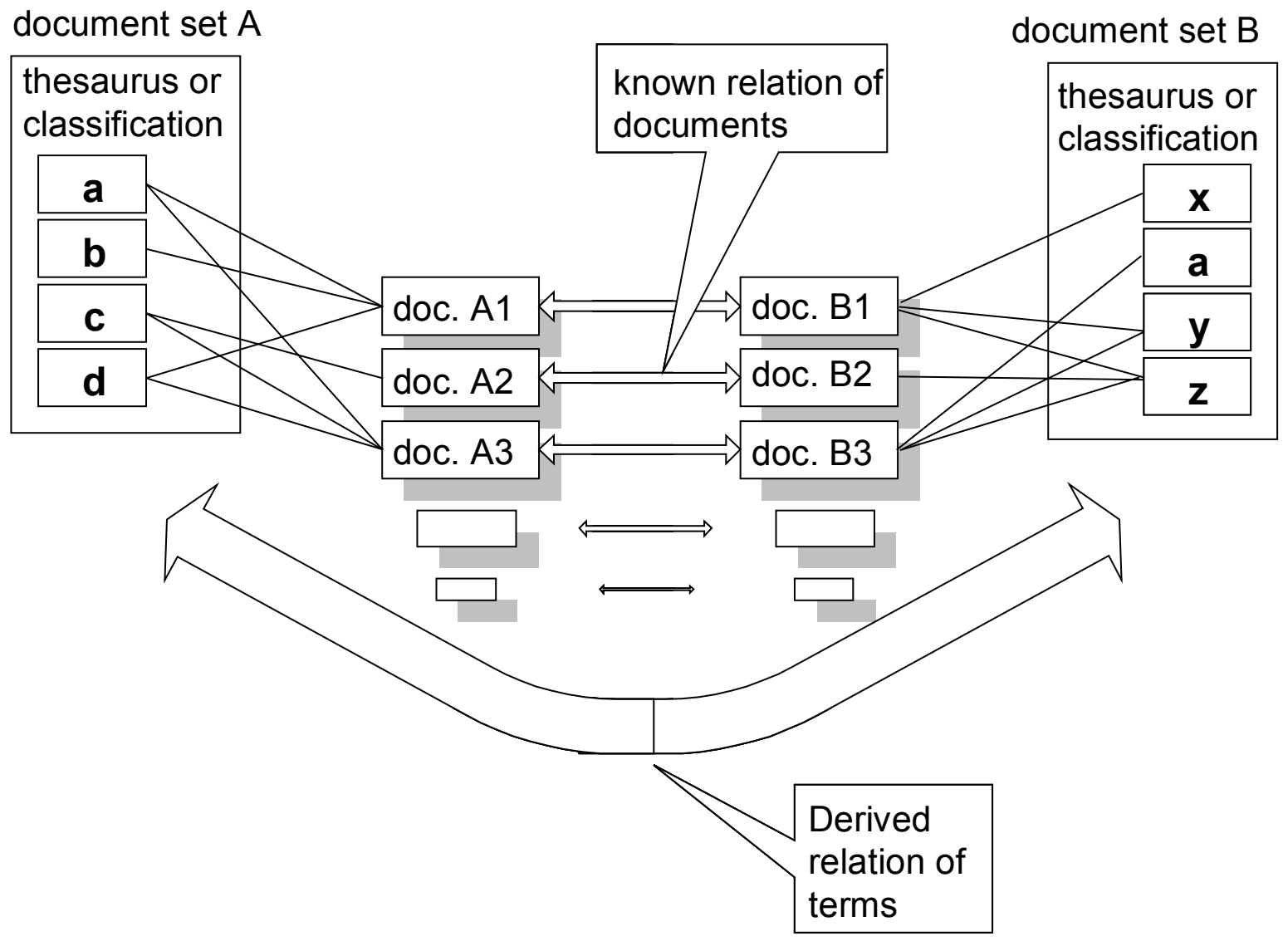

**Figure 8: Parallel Corpus**

After the creation of the parallel corpus, the indexation terms of document *Ax* from data set *A* can be brought into relation with the indexation terms of the same document *Bx* from data set *B*. Lets consider the following document as an example:

***'Gysi, Jutta: Familienleben in der DDR, zum Alltag von Familien mit Kindern, Akademie Verlag Berlin, 1989, ISBN 3-05-000771-0.'***

It has been indexed by the library of the University of Cologne with the terms

***'Deutschland <DDR>'*** (Germany <GDR>) and ***'Familie'*** (family)

from the SWD (Subject Authority File of the German national library, Die Deutsche Bibliothek). Whereas the same document has been indexed by the IZ with the terms

***'Arbeitsteilung'*** (division of labor), ***'Ehe'*** (marriage), ***'Familie'*** (family), ***'DDR'*** (GDR) and ***'Partnerschaft'*** (partnership) from the Thesaurus for the Social Sciences (edited by the IZ).

Now the different terms can be brought into relation. This is done by relating every term from indexation A to every term of indexation B, e.g. the SWD term *'Deutschland <DDR>'* will be related to the social science thesaurus term *'DDR'*. Thus, this example document will result in ten relations (two SWD terms multiplied by five social science thesaurus terms). Of course not all of these relations make sense. The few really useful ones will be filtered out by the co-occurrence analysis.

## 4.2 1 x N Relations

Basically, statistical co-occurrence analysis is an analysis of the number of occurrences of a relation between two terms in a corpus. Therefore the corpus as a whole and not the individual documents are of interest. For example, if the relation *'Deutschland <DDR>'* → *'DDR'* appears not in just one document, but in quite a few, this seams to be a useful transfer relation. The decision whether or not a relation makes sense is quite difficult. Usually it is not based on the pure number of occurrences, but is computed out of different factors by different mathematical models (see section 4.4). The underlying data sets determine which model is best used with which factors, and with which thresholds. The precision of each method depends on the size of the corpus. The larger the corpus, the better their precision.

Compared to the intellectual cross-concordances, statistical transfer relations have just one type of relation: the terms are (somehow) related. A qualification, for instance that they are synonyms or one is a broader term of the other, is not possible. Instead of this, the closeness of the terms can be determined very precisely; usually it is given on a scale between zero and one (see Table 1).

| | *DDR* | *Familie* | *Ehe* | *Partnerschaft* |
|---|---|---|---|---|
| ***Deutschland <DDR>*** | 0.93 | 0.40 | 0.10 | 0 |
| ***Familie*** | 0.50 | 0.88 | 0.18 | 0.09 |
| ***Zweierbeziehung*** | 0 | 0.40 | 0.40 | 0.50 |

**Table 1: Examples of statistical closeness of terms**

Nevertheless, taking a closer look at statistical transfer relations, a number of one to one relations, stating that one term relates to exactly one other term, will be found. Those terms usually are synonyms.

The second group of relations is made up of one to many relations (1x N). Those relations usually have three origins:

- hierarchical information; the term on the left hand side is a broader term of those on the right hand side.
- ambiguous term on the left hand side; the terms on the right hand side represent its possible meanings.
- nominal composition on the left hand side; the terms on the right hand side are the single parts of the composition.

Nominal compositions are quite common in the German language. Considering the example from above – the transfer relations between the SWD and the social science thesaurus – we will find a lot of relations decomposing a composed noun into the terms the composition is made of. This happens, because the SWD is pre-coordinated whereas the Thesaurus for the Social Sciences is post-coordinated. In pre-coordination the thesaurus already contains the compound forms (e.g. '*Jugendarbeitslosigkeit* '/youth unemployment), whereas in post-coordination the thesaurus contains only the different parts of the composition (e.g. '*Jugendlicher*' and '*Arbeitslosigkeit*'), and the actual combining has to be made at retrieval time. Therefore every compound noun should result in a one to (usually) two term relation.

The different semantic background of hierarchical, ambiguity and nominal composition based relations causes a problem in the creation of those relations. The difficulties arise from the different connectors that the different types require. The hierarchical and ambiguity based terms have to be connected by OR, whereas the nominal composition based ones have to be connected by AND. The decision when to use which operator is quite complex, and cannot be solved by pure statistics. Therefore we have chosen to use a workaround as follows.

On the right hand side of the relation both single terms and combinations of those terms are allowed. Therefore every possible permutation of the terms of a document on the right hand side is computed. Regarding our example from above the five terms from the social science thesaurus would result in 25 new term combinations (e.g. '*Familie* ', '*Familie* AND *Ehe*', '*Familie* AND *Ehe* AND *Partnerschaft*', …). This leads to 50 instead of 10 resulting possible transfer relations. Again the assumption is that only those relations that make sense will remain after the computation of the statistical closeness.

Therefore the remaining terms/term combinations can be combined with the connector OR. In the case of nominal composition only one of the permutated term combinations should remain.

Clearly this problem is not finally solved, and maybe an additional linguistic post-processing of the statistically generated transfer relations is needed.

The situation for classification is less complex. Nominal composition does not exist, and therefore the connecting operator to be used is clearly OR.

## 4.3 N x M Relations

Besides the common one-to-one / one-to-many relations, many-to-many transfer relations (N x M) are possible. Regarding Boolean retrieval, those relations are quite rare and can be regarded as a more theoretical problem. Nevertheless the discussion of the two problem domains – creating and applying those relations – gives a deeper insight of the overall problem domain of statistical transfer relations.

Considering the creation of N x M relations it becomes clear, that in general this is the case for all transfer relations. Groups of terms – a single term is still a group – have to be related to groups of other terms. The problem arises from the different semantics of the Boolean operators OR and AND which will connect the N and M terms. To find every useful combination of terms and operators, all possible permutations must be generated and evaluated by the statistical analysis. This process is quite complex and resource intensive. Besides the exponential growth of required recursions, it becomes very difficult to find the appropriate threshold values for the statistical methods (see next section).

Once the transfer relations have been computed the next problem arises; how to apply them? Because of a possible embedding of the source terms of a relation, this problem is much more complex then the one with one-term translations. Lets consider the following example.

The transfer relation

'*adolescent*' AND '*criminality*' → '*youth criminality*'

should be applied to the query

'*criminality*' AND ('*youth*' OR '*adolescent*').

This cannot be done by direct pattern matching. A normalizing process – including alphabetic ordering of the terms – has to be applied first. This will result in the equivalent query

('*criminality*' AND '*youth*') OR ('*adolescent*' AND '*criminality*').

Now the query can be transferred – via pattern matching – into
('*criminality*' AND '*youth*') OR '*youth criminality*'

Again this process is quite time and resource consuming. In particular, the time constraints are problematic. Matching has to be done during the retrieval process, and therefore adds to the over all response time of the information system.

## 4.4 Models

During the last two decades various mathematical models have been suggested for the analysis of the co-occurrence of terms. Their application was on automatic indexing mainly (Biebricher et.al. 1988; Ferber 1997), but they have been used for term expansion and clustering also (Grivel/ Mutschke/ Polanco 1995). The conditional probability and the equivalence index achieved some of the best results.

Basically the **conditional probability** is the probability of the occurrence of a term *b* from thesaurus *B*, if the same document has been indexed by term *a* from thesaurus A. Thus stating that there is a probable transfer relation between term *a* and term *b*. Formula 1 shows the according computations.

$$P(a \rightarrow b) = \frac{P(a \,\&\, b)}{P(b)} = \frac{\frac{C_{a\&b}}{C_{all}}}{\frac{C_b}{C_{all}}} = \frac{C_{a\&b}}{C_b}$$

- P(a & b) is the probability of term a and term b occurring together in one document
- P(b) is the probability of term b occurring in one document
- $C_x$ is the number of documents with term x assigned
- $C_{all}$ is the number of all documents in the parallel corpus

**Formula 1: Conditional Probability**

The **equivalence index** is a measure of the similarity of two terms. Therefore the resulting relations are none directed ones. The advantage of this equality is that very closely related terms can be filtered out easily (they will have a very high equivalence index). The disadvantage is that loosely related terms are very hard to find, because of their very low equivalence index.

$$P(a \leftrightarrow b) = E_{ab} = \frac{C_{ab}^{\;2}}{C_a * C_b}$$

**Formula 2: Equivalence Index**

Those mathematical models are quite simple, but their advantage is, that they are easy to scale, which allows a wide range of applications. The next section describes a tool for modifying the summation algorithm and to find the appropriate threshold values.

## 4.5 Jester

During the project ELVIRA, a tool for generating statistical correlation relations based on parallel corpora was implemented. JESTER (the Java Environment for Statistical Transformations) is a general workbench that allows for the interactive selection of parameters for optimizing the transfer relation between a pair of classification systems. JESTER also employs a number of heuristics for the elimination of systematical errors, introduced by the simulation of an actual parallel corpus as described in section 5.

In particular, the graphical representation of the term-document frequencies (cf. Figure 9) permits the eliminations of documents and/or terms from the following steps, based on their occurrence. In the case of a simulation of a parallel corpus (described in section 5), some documents got too many terms assigned. This happens, when the probabilistic search engine erroneously re-turns the same document on almost every query because some domain specific phrase appears verbatim.

On the other hand, terms that appear in too many documents have almost no semantic discrimination value for the document collection at hand. Terms that only appear in very few documents don't offer enough evidence for the statistical correlation analysis. Eliminating both of these term groups from the sample generally improves the quality of the result, not only in simulated parallel corpora, but also when index terms have been assigned manually.

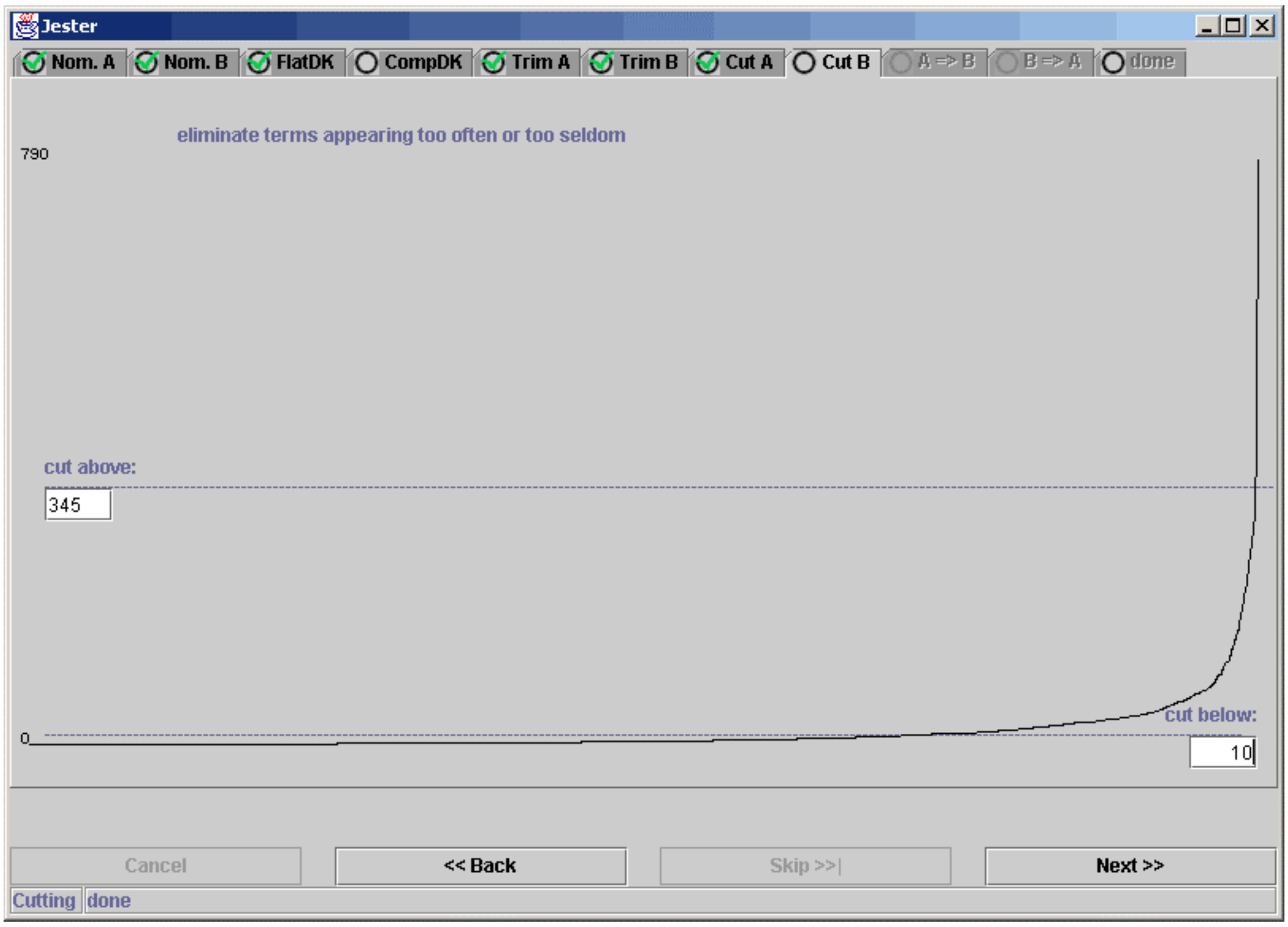

**Figure 9: Interactive elimination of high and low frequent terms in a graphical representation**

The following correlation technique is based on assigning each possible pair of terms a weight in the range [0, 1], calculated from the individual weights of the document-term relations and the number of occurrences. These weights can further be modified (lowered) by multiplying them with the conditional probability for each direction of transformation. Multiplying by both of these measures is equivalent to the incorporation of the equivalence index, introduced in section 4.4.

The heuristic for favoring root- or leaf terms (discussed in section 4.6) can be used to gradually lower the weight of term-term relations that do not meet the criteria of specificity or generality. Different cut-off thresholds can be applied to these modified weights, as well as to the basic conditional probabilities, in order to eliminate the unwanted correlations. All of these parameters can be manipulated simultaneously in one central configuration dialog (cf. Figure 10).

While manipulating these parameters, a preview for the resulting translations of a small selection of example terms can be displayed and updated in real time (cf. Figure 11). This immediately shows the effects and allows for the interactive exploration of the interactions and dependencies of the multiple parameters in use. Once an optimized parameter setting for the transformation direction has been determined, the correlation process can be applied to all term pairs and the result, and is then stored in a file.

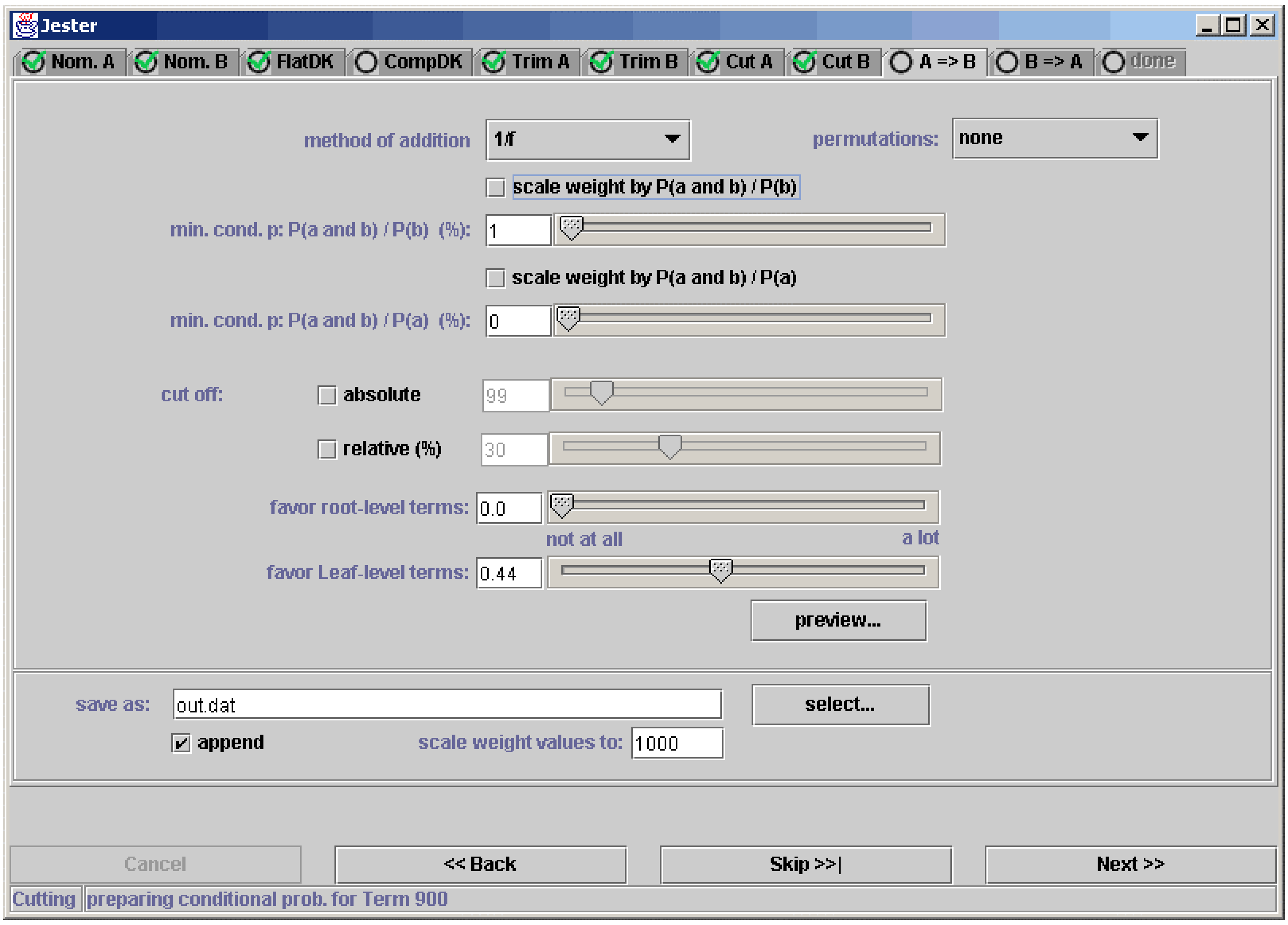

**Figure 10: Parameter specification in JESTER**

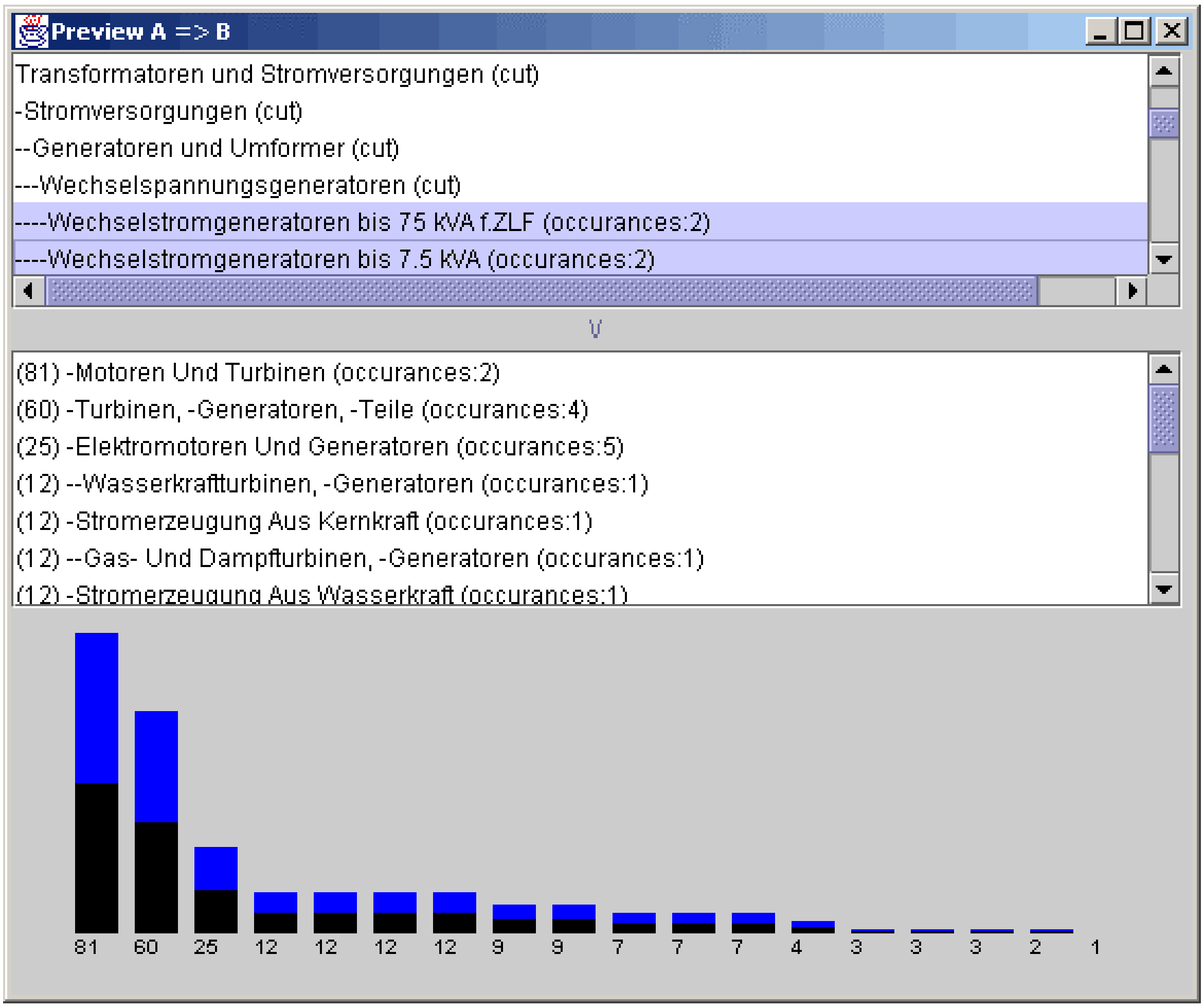

**Figure 11: Dynamic preview of the results for the current parameters**

## 4.6 Use of Keyword Hierarchy Information for Fine Grained Control of the Transformation Result

During the course of the ELVIRA project, it became obvious that, depending on the intended use of the transformation process, different flavors of correlation were desirable for each pair of classifications and even for each direction of query transformation. Depending on the documentary practice in assigning keywords within each document collection and on the structure of the keyword system in use, it is sometimes desirable to perform a broadening query transformation, resulting in a query with common higher level concepts. For other document collections, a transformation containing only the most specific of the corresponding terms is better suited. Also, it is important to consider the way the user interacts with the system. If the user is meant to review or modify the new query (whose vocabulary he is at least passively familiar with), a higher level of abstraction should be used than when the transformed query is to be executed without further user interaction.

The common parameters which could be identified within these different approaches could be mapped to the positions of terms within the hierarchy of the destination classification system of each transformation direction. For parametric control, the root and leaf distance of each term within its classification tree structure was computed. The root distance is defined as the number of child-parent relations to be traversed within the hierarchy, until the root node of the tree is reached. Likewise the leaf distance is defined as the (minimal) number of parent-child relations to be traversed until a leaf (an entry without further children) in the tree structure is reached. Using these metrics, a formula for modification of the weight information was developed that allows for the favoring of correlations to terms which are either close to the root of the destination classification tree (more general terms) or close to leafs (more specific terms).

$$W'_{(termA,termB)} = W_{(termA,termB)} \left( \frac{1}{leafDist_B} \right)^{leafFavor} \left( \frac{1}{rootDist_B} \right)^{rootFavor}$$

**Formula 3: Modification of weight, depending on the hierarchical position of the destination term**

# 5 Parallel Corpora Simulation

## 5.1 Parallel Corpora Simulation with Nomenclatures and Classifications

### 5.1.1 Problem Description

In some domains or project contexts, the goal of coupling any number of semantically compatible classifications lead to the aim for a (semi-) automated concordance generator. In the project ELVIRA for example, a concordance to a domain specific meta thesaurus was created manually for parts of the data. The number of nomenclatures and classifications in use and the high fluctuation made it obvious, however, that this approach could only be employed for a core set of data. Less central data pools could not be integrated timely in this fashion.

The basic data for automated concordance generation (cf. Figure 12) consisted of a number of hierarchical nomenclatures for the factual data and a sample body of text data (BfAI) which was already indexed manually according to a controlled set of (also hierarchically structured) keywords. Parts of the nomenclatures were furthermore accompanied by short, explanatory text fragments. An actual parallel corpus was not available. The fact data itself was not used in the creation of the correlations.

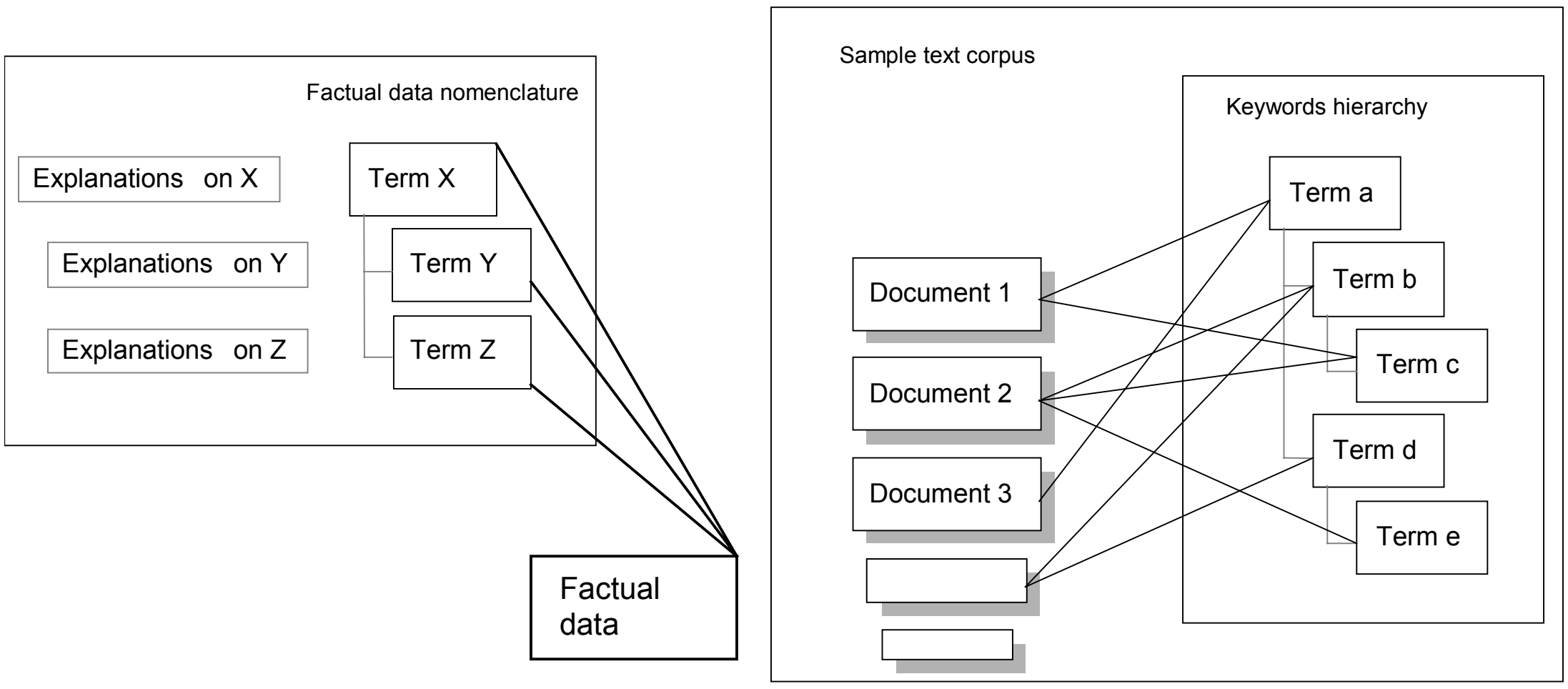

**Figure 12: Base data for parallel corpora simulation in the project ELVIRA**

Since the fact nomenclatures frequently change, manual re-indexing of the text corpus for creating an actual parallel corpus did not seem a valid approach. The indexing step would have to be repeated after each major change of the nomenclatures or inclusion of a new one. Instead, we developed an approach that uses a probabilistic search engine to simulate this manual indexing step and assign nomenclature terms to documents from the text corpus automatically, based on the available rudimentary information.

### 5.1.2 Solution

In order to simulate a parallel corpus, the sample text body was indexed by a probabilistic search engine (Fulcrum). In probabilistic indexing, each document, as well as the query, is represented by a vector. The search process returns those documents, whose vector matches the query vector most closely, together with a correlation measure. This measure shows, how close each returned document corresponds to the query.

In an automated preparation step, each term of the fact nomenclatures was augmented with the provided explanatory text fragments, some domain specific abbreviations were expanded and common terms and domain specific stop words were eliminated. In nomenclature hierarchies representing sub/super concepts (as opposed to part/whole relations), elements were further augmented with the terms of their direct super concept to provide context.

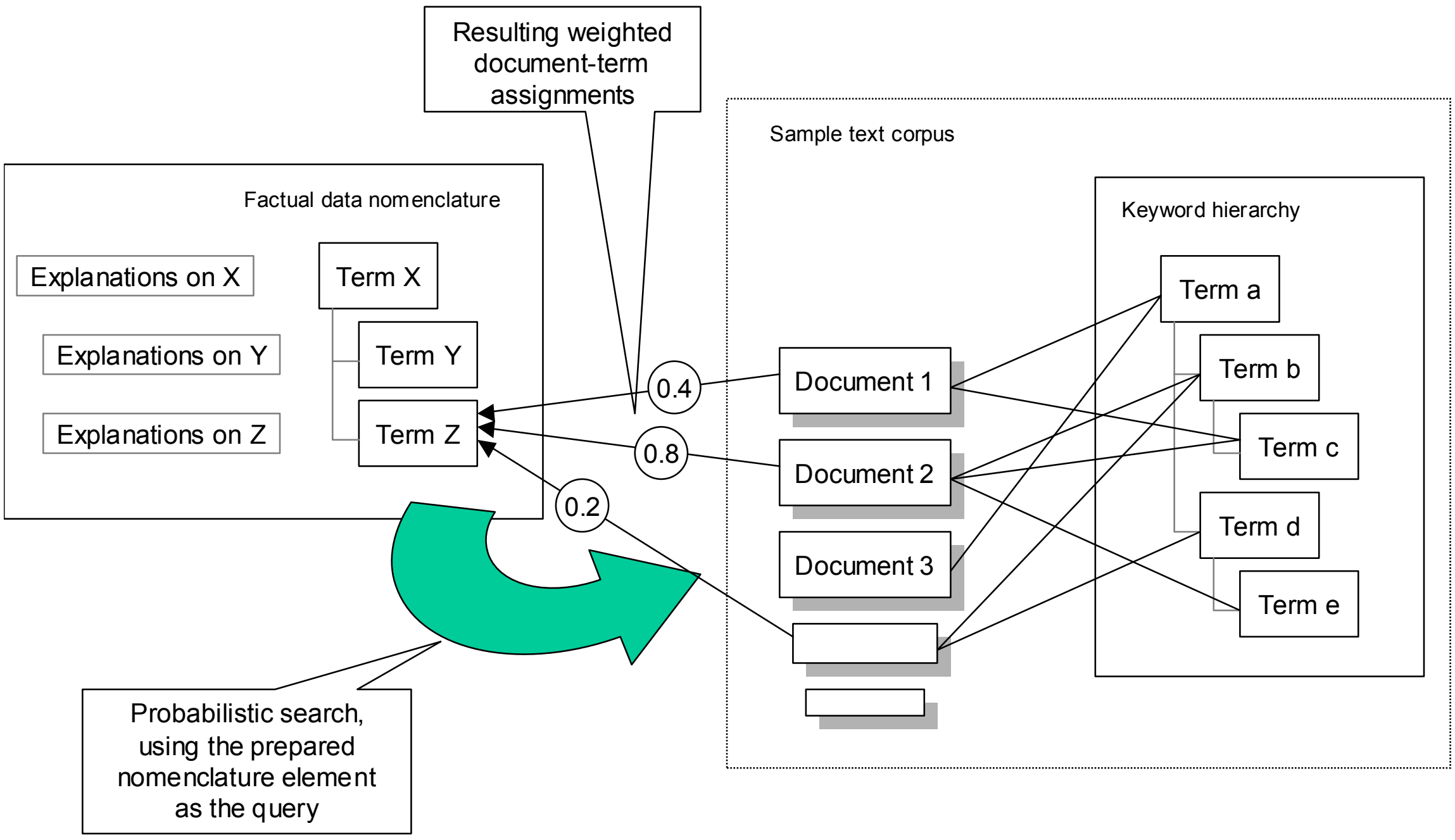

**Figure 13: Simulation of a parallel corpus by use of a probabilistic search engine**

Each of the resulting phrase collections was then used as a query to the probabilistic search engine (cf. Figure 13).

The weight information returned with the resulting document set was made comparable between queries by application of some simple, search engine specific heuristic. For each document in this result set, the nomenclature term used for this query was stored, together with the equalized weight information of the document within this specific search. The term–document relations whose weight exceeded a certain minimum weight threshold were now used like keyword assignments, resulting from manual indexing.

To further use the information returned from the search engine, a weight measure was introduced into the following correlation process. We defined that manually assigned keywords were “pure” assignments and assumed a weight of 1 on a scale from 0 to 1. The term-document relations resulting from our simulation were interpreted to be only partially matching keywords and were assigned a smaller value, based on the equalized weight returned by the search engine.

The measures for document relations, described in section 4.4 had to be adapted to deal with these partial term assignments. The conditional probability definition is based on the terms, specifically the term pairs, probability. This probability was previously defined to be the number of documents containing the terms in question, divided by the total number of documents on any given document set. The easiest approach to introducing the weight into this calculation is to replace the number of documents containing the terms with the sum of the weights of the terms over all documents. For the simple case, where only manual document assignments of weight 1 are present, the formulas are equivalent. Also, if all assignments of any given term to a number of documents have a lower but equal weight, the computed value is decreased by the same factor.

This approach however does not allow us to discriminate between many low weighted occurrences vs. few, high weighted occurrences. Therefore, a number of experimental heuristics which modify the individual term weights according to the overall term frequencies was implemented. Though these yield very promising results for individual pairs of classifications, no general solution could be identified. Instead, the selection of the summation method remains an optimization parameter of the statistical workbench that is used to generate the transfer relations (cf. section 4.5).

## 5.2 Parallel Corpora Simulation with Keywords and Full-Text Terms

In dealing with the World Wide Web we are concerned with the frequency with which Web pages, as well as documents accessible via the Web, are not indexed by a specific (given) thesaurus or classification system. The only terms we can rely on are the full-text terms supplied by a full-text indexing machine.

Taking into account that a user might start his search with controlled vocabulary obtained from a thesaurus, relevant internet documents should be retrieved as well as well-indexed documents stored in a database. In order to provide this, we have to realize a transfer from classification terms, thesaurus terms respectively, to full-text terms, and vice versa. As long as we can not fall back to any standards of classifying internet documents, we have to use a weaker strategy of combining keywords and full-text terms. Note that intel-lectual indexing would result in enormous costs. This weaker strategy is simulating parallel corpora for supplying semantic relations between keywords and internet full-text terms.

In order to provide a simulated parallel corpus, we have, first of all, to simulate intellectual keyword indexing on the basis of a given thesaurus. Simulating intellectual indexing implies that a method is used that produces *vague* keyword-document-relationships, i.e. unlike intellectual indexing, where each assignment is (usually) un-weighted, weighted by 1 respectively, simulated keyword-document-ties are weighted on a [0,1]-scale. This yields a situation as indicated in Figure 14: Unlike the situation in public databases (like the German social science literature database SOLIS), where we have exact assignments of keywords and documents, we produce vague keyword indexing as well as vague full-term indexing.

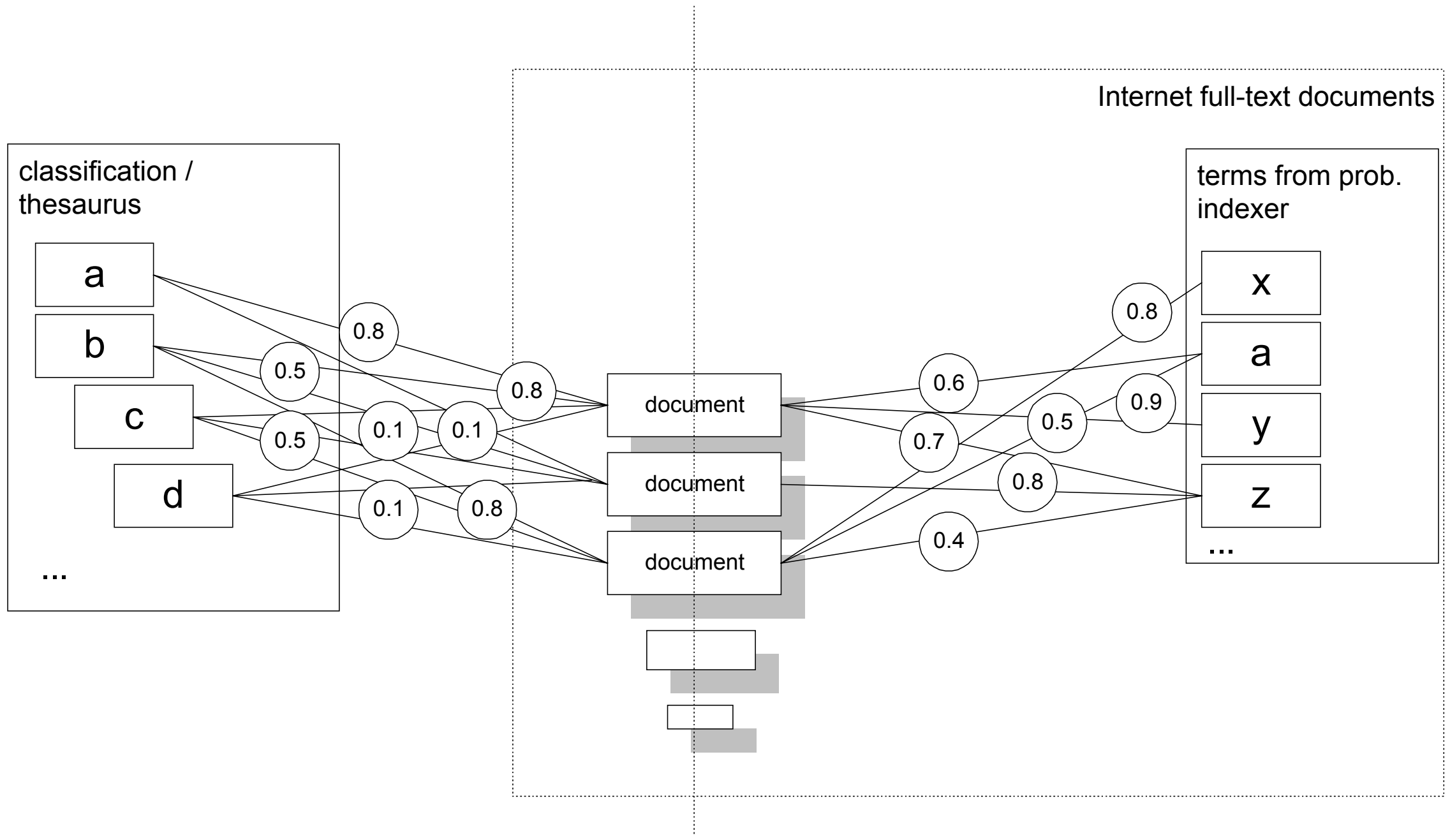

**Figure 14: Parallel Corpus Simulation with vague keywords and full-text terms**

In the following parallel corpora simulation via vague keyword and full-text term assignments is described for the CARMEN test corpus. The CARMEN test corpus is a collection of about 6.000 social science internet documents that have no keyword assignments.

### 5.2.1 Assignment of Keywords and Full-Text Terms to Internet Documents of the CARMEN Corpus

For the assignment of controlled vocabulary (keywords) to non-classified internet documents a given thesaurus is used, i.e., in the case of the CARMEN corpus, the thesaurus for Social Sciences (IZ 1999). As a basic method assigning thesaurus keywords to internet documents we consider each single keyword of the thesaurus as a query that is „fired" against a full-text indexing and retrieval machine maintaining the CARMEN corpus. The full-text retrieval engine Fulcrum SearchServer 3.7 is used to provide ranking for values the documents in the corpus according to their similarity to the query vector. Each document in the result set that has a ranking value greater then a certain minimum threshold is then indexed by the keyword requested. The ranking values supplied by Fulcrum are considered as the weights for the keyword assignments.

This basic method is not sufficient because it does not consider the relevance of a keyword for the document retrieved. It indicates only that a document contains a keyword but not whether the document is well-described by it as regards the documents contents. Therefore, some topical context information regarding each individual keyword has to take into account. For this purpose a further parallel corpus is build up containing SOLIS documents that thematically correspond to the Carmen corpus. Each SOLIS document is a structured record consisting of a set of bibliographic fields, like title and abstract, and a set of (usually ten) keywords that have been intellectually assigned to the document. Since title and abstract are not full text, note that terms appearing in those fields are called *free-text terms* in the following. Useful context information for each keyword is provided by evaluating the co-occurrence of keywords and free-text. This is done by tokenising the title and abstract string, eliminating stop words, and stemming the terms using a Porter stemming algorithm for German. On the basis of the final term list three types of co-word-relationships between the initial keywords and the terms in the term list can be determined: keyword/keyword-ties, free-term/free-term-ties, as well as keyword/free-term-ties. For the calculation of the strength of these relationships the *equivalence* index is taken, which indicates the semantic similarity of co-words. To reduce the set of tuples generated to a meaningful set, only co-words are used that do not fall under a certain threshold. The initial query vector is then enriched by keywords and free-text terms co-occurring with the initial keyword in corresponding SOLIS documents. This procedure yields an extended query vector that is now "fired" against the CARMEN corpus via Fulcrum. As ranking algorithm *term count* is used emphasizing documents matching most of the terms in the query vector. By this way it is ensured that only documents dealing with the initial keyword in a thematically appropriate manner are selected. The most relevant documents are then indexed by the initial keyword.

Full-text terms are obtained by tokenising the full-text of an internet document, eliminating stop words, and stemming the remaining terms using a Porter stemming algorithm for German. For weighting the terms the *inverse document frequency* (Salton 1987) is used. The full-text is then indexed by full-text terms having a weight greater than a certain minimum threshold.

# 6 Neural Networks

Neural networks are a computing method which belongs to the paradigm of soft computing. The features of soft computing as opposed to exact computing all apply to neural networks (cf. Zadeh 1994).

Neural networks use the architecture of the human brain as a role model. They consist of a large number of simple distributed processing units all working strictly local and communicating along connections between them. Only the overall pattern of activation can be interpreted meaningfully whereas the contribution of a single calculation to the whole process can hardly be identified.

Each neuron has a numerical activation strength storing its current state. In an artificial neural network this activation can change at each time interval. The connections between the neurons are directed and associated with weights influencing the activation within the network. These weights are typically modified during learning processes which is again an analogy to the biological brain. The synapses between neurons can change in order to let more information flow along them. Figure 3 shows these basic processes of a unit or artificial neuron.

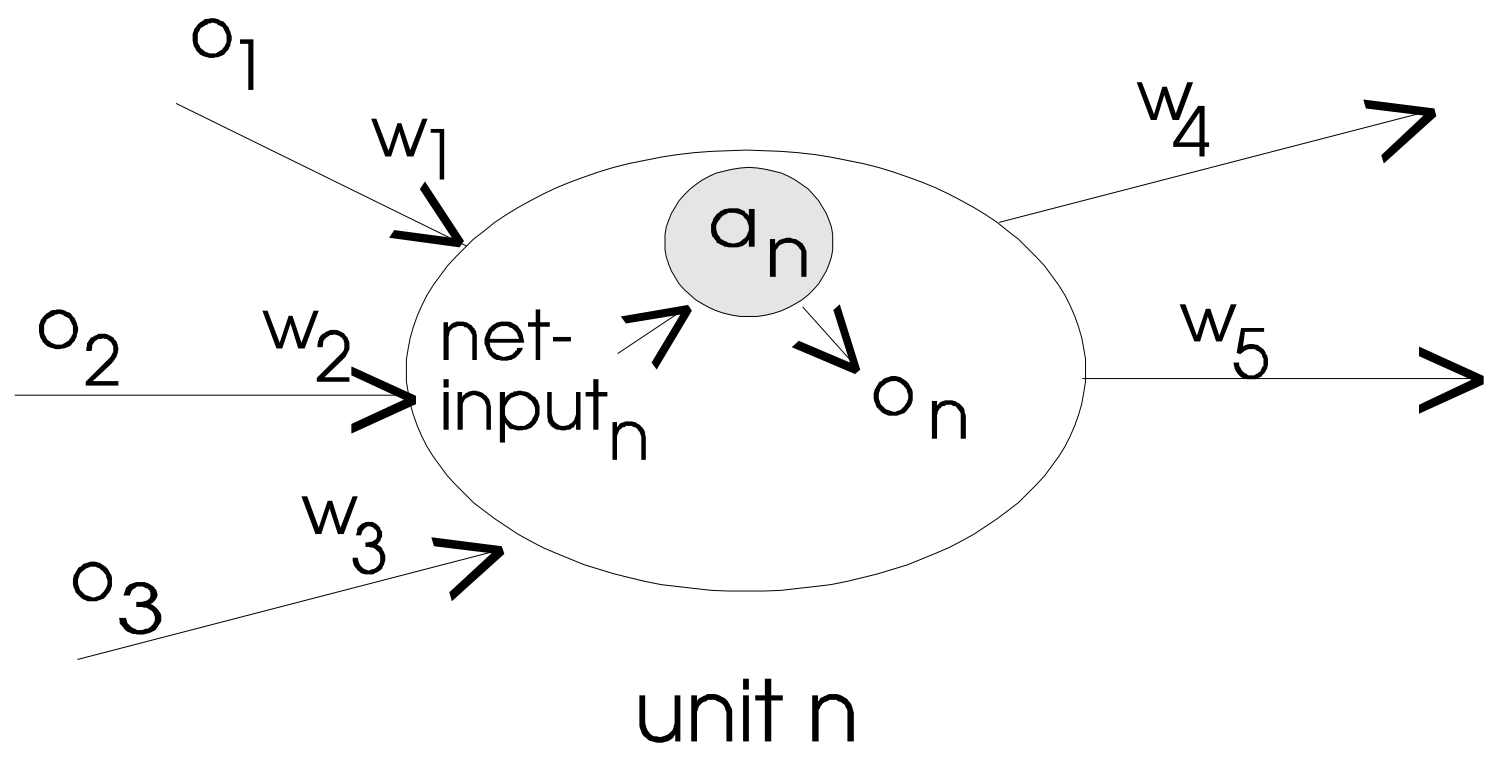

| | |
|---|---|
| *$o_i$:* | *output of unit i* |
| *w :* | *connection weight* |
| *net-input$_n$:* | *input for unit n* |
| *$a_n$:* | *activation of unit n* |

**Figure 3: Basic operations of an artificial neuron (cf. Dorffner 1991:17)**

Each neuron calculates its own activation based on the input received from incoming connections. The simplest rule for determining the activation level is a threshold function which sets the activation to the maximum when the input is at least the threshold and sets it to zero when the input is lower than the threshold. Often smoothed differentiable activation functions are used. The activation of a unit triggers the output. The output travels along all outgoing connections to other units. The strength of the connection weights determines how much of the output activation actually reaches the other units. Most often, the output is multiplied with the connection weight. In the next time interval, this activation becomes input for the other units. The spreading of activation over many intervals is one of the most typical features of neural networks.

Initially some of the units receive activation from outside. These stimuli serve as input. After the processing, another set of units serves as output and presents the reaction of the network to the outside, which can be a user or another computer program.

Many different models of neural networks are based on these principles. They differ in their connections matrices, the activation functions and especially in their learning rules.

The parallel and distributed processing paradigm leads to highly tolerant systems. Neural networks can learn from existing data even when humans find it difficult to identify rules.

*Neural or connectionist networks have successfully been applied to a large range of problems like optical character recognition, control of autonomous vehicles and time series forecasting. Neural networks seem to be well suited for information retrieval tasks and have attracted considerable research in that area as well (cf. Mandl 2000).*

A number of different types of networks have been introduced to information retrieval. For example associative memories like the Hopfield network, which are powerful error-tolerant storage and retrieval tools (cf. Bentz et al. 1998), or Kohonen self organizing maps (SOM), which can be used for unsupervised clustering (cf. Kohonen 1997). But the backpropagation model still seems to be the most popular.

## 6.1 The Backpropagation Model

The backpropagation network is a powerful supervised learning method and can be regarded as the most popular neural network architecture. The connections run from the input layer over one or more hidden layers towards a layer of output units. The learning algorithm of the backpropagation network learns a mapping between input and output space based on examples.

The input vector is propagated into the network that calculates the output vector. In a training phase, examples for the desired mapping are presented enabling the network to compare the actual output and the desired output. The error is used to tune the connection strengths of the network such that it is closer to the teaching output and that it can better approximate the desired function. After training the network, its generalization capabilities need to be tested with another data set (cf. Haykin 1994).

The adaptation of the connection weight between two units is calculated with the delta rule, which considers the error as difference between target or teacher and actual activation and the output of the starting unit as a measure of its contribution to the error (cf. Mitchell 1997:89):

*Delta-Rule*

$$\Delta w_{ij} = \eta \, output_i \, (teacher_j - activation_j)$$

$\eta$ *learning rate*

The hidden units increase the computing abilities of the backpropagation network and enable it to learn complex functions such as non linear separable problems (Mitchell 1997:105). Backpropagation like other supervised learning methods can be regarded as the approximation of an unknown function. The whole process of backpropagation learning is illustrated in Figure 4.

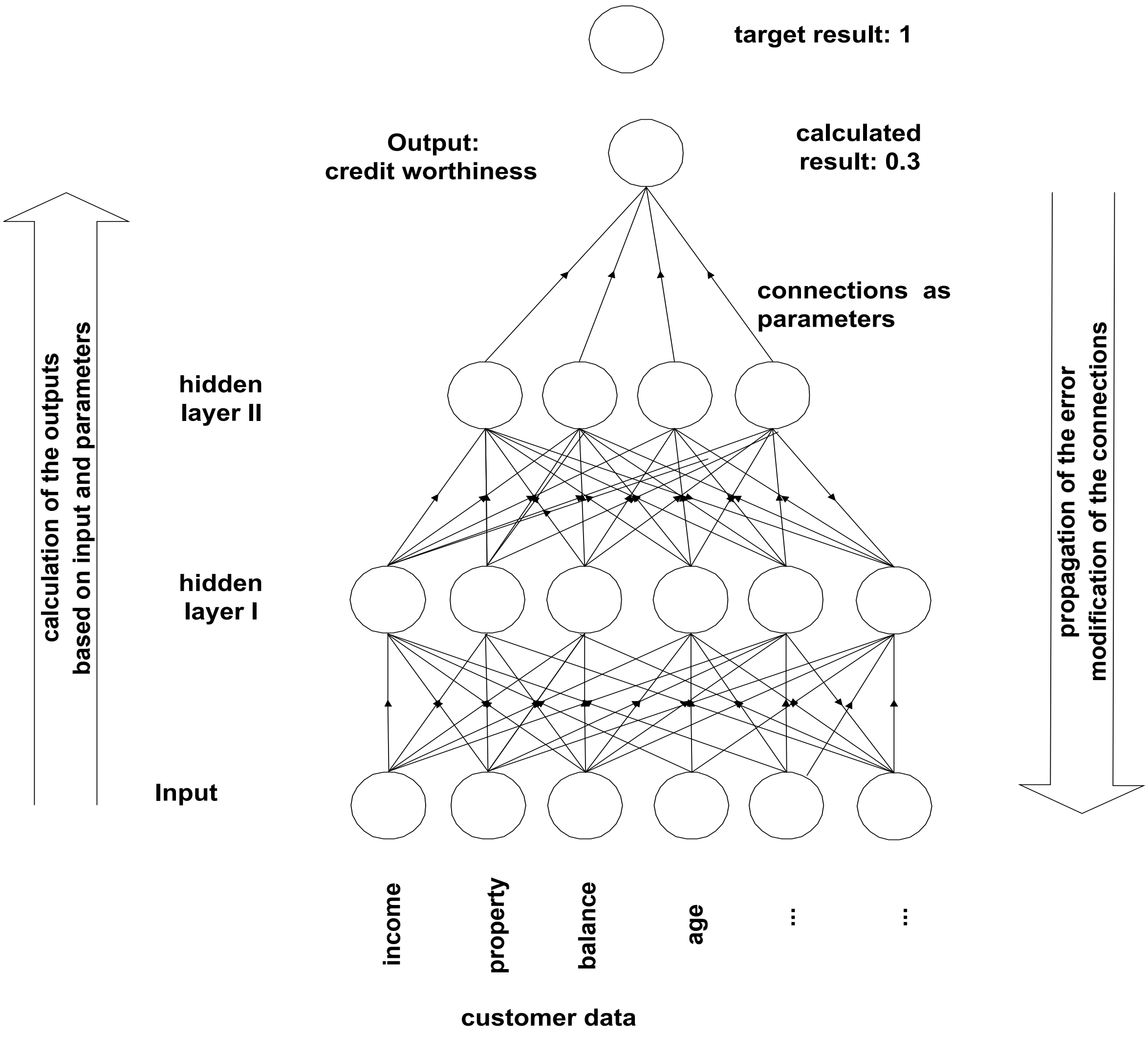

**Figure 4: Example for the application of the backpropagation network**

## 6.2 Neural Networks and Information Retrieval

The soft computing paradigm of neural networks seems to be well suited for Information Retrieval tasks. This particular field has attracted considerable research; however, the search for an appropriate architecture has proved to be difficult. Current systems can be grouped into four categories.

The spreading activation networks have performed well in experiments within the TREC conference (e.g. Boughanem and Soulé-Dupuy 1998). However, a closer look at the models reveals that they very much resemble the traditional vector space model of information retrieval. Mothe 1994 presents a theoretical and empirical proof that after the first step of activation propagation, the models are indeed identical. Thus, spreading activation networks are not a new model for information retrieval.

Several researchers have implemented Information Retrieval systems based on the Kohonen Self-Organizing Map (SOM), as a neural network model for unsupervised classification. Implementations for large collections can be tested on the internet (Chen et al. 1996, Kohonen 1998).

Although the backpropagation algorithm is one of the most powerful and most often used neural network models, it has not been applied to information retrieval very often to date.

The capability of backpropagation for tolerant and intuitive processing seems to make it a promising approach for information retrieval and the vague nature of human relevance judgments involved. The approach we have experimented with uses a backpropagation network to create transfer relations between groups of terms. Therefore the network is trained with documents of a parallel corpus. The terms of the content analysis of a document from the source thesaurus make up the input vector, whereas the terms for the target thesaurus build the output vector. Trained with a lot of documents, the network learns to transfer a given vector/group of terms to another group of terms. The application of a such trained net is described in section 7.4.

# 7 Integration of Transfer Relations into Retrieval

## 7.1 General Model of Integration

Once the transfer relations have been realized, the question remains how to incorporate them into information retrieval systems. As a query term manipulating methodology they have to be placed somewhere between the user inter-face and the databases. For a better understanding of the transformation process, we should take a closer look on the architecture of a distributed information retrieval system. Transfer relations, or – to be precise – transfer modules, become necessary, if data sets have to be combined, which are indexed by different content analysis systems. Usually those data sets reside in different databases. This distribution of data demands for a three-tier-architecture, consisting of user interface, coordination layer and databases. In a classical coordination layer the user query is simply distributed – unchanged – to the different databases and the results are combined to an overall result set. Most of the meta search engines in the WWW work this way.

But this procedure is not applicable to data with heterogeneous indexing systems. To send one and the same query to all databases would lead to a lot of zero results in the connected databases. This is due to the fact that e.g. a queried classification is available in only one database, but not in the others. At this point the transfer relations come into play. Through their ability to transform controlled vocabulary, they are able to adapt the users query to the different requirements of the database. Therefore the query the user has issued will be transformed into different queries fitting the different content description systems (cf. Figure 17).

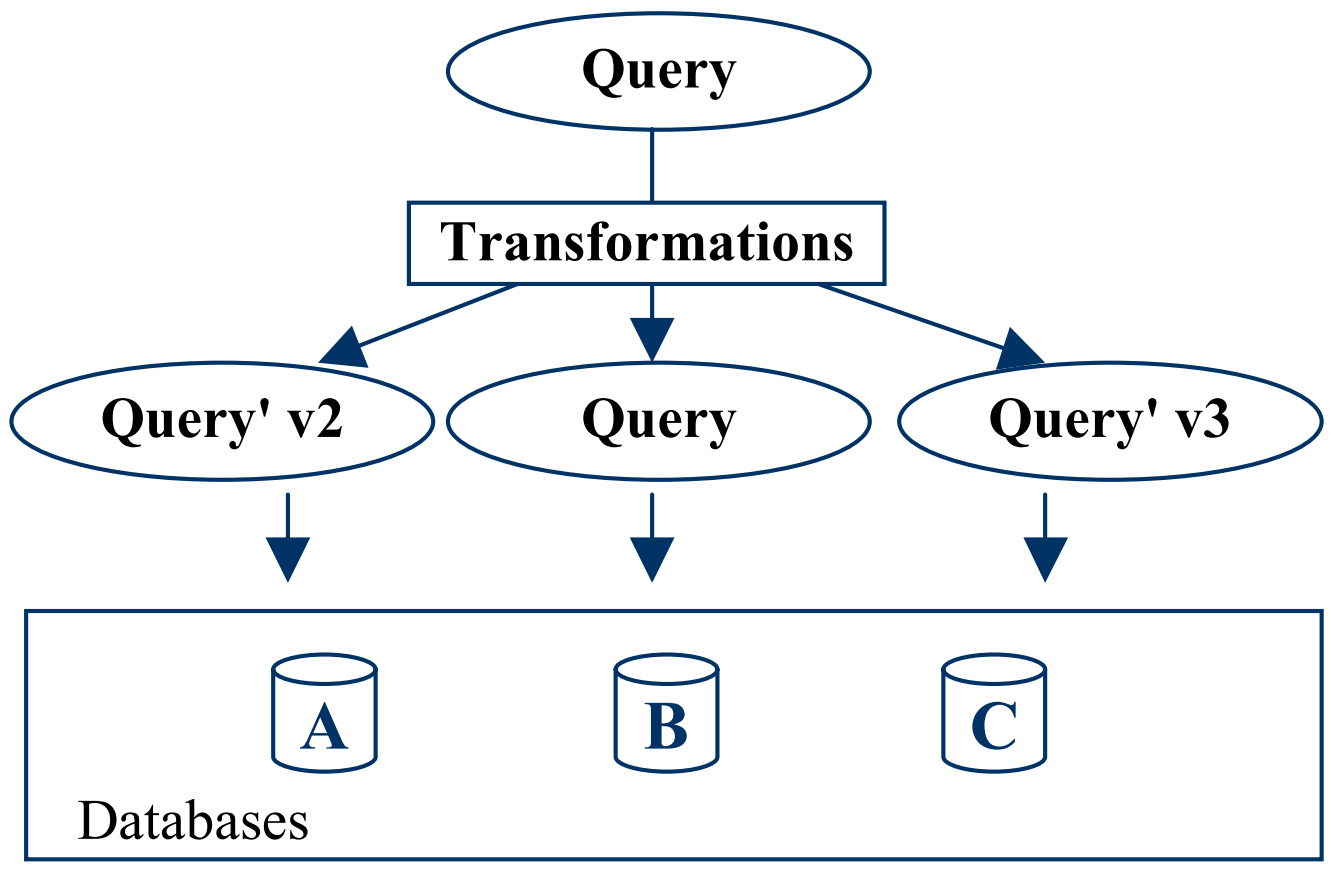

**Figure 17: Query transformation process**

During the actual transformation the relevant part of the users query (e.g. the classification terms from system *A*) is separated from the other parts. The terms of this separated part act as the input for the different transformation modules. After the transformation, the resulting output forms the new, transformed query part. This new part consists of terms from system *B*, and is combined with the rest of the users query (e.g. author/title query) to form the new, transformed query. Afterwards the query is send to the corresponding database.

## 7.2 Transfer Use in CARMEN

The retrieval system *HyRex*[13] of the project CARMEN is part of another package implemented at the University of Dortmund.[14] This search engine uses transfer services, which run at the IZ. Relevant parts of the complete query (e.g. author is no transferable query type) are sent to the transfer service as XIRQL (XML *Information Retrieval* Query Language) statements by http request and answered with a new transferred XIRQL statement that represents the transferred query (cf. Figure 18).

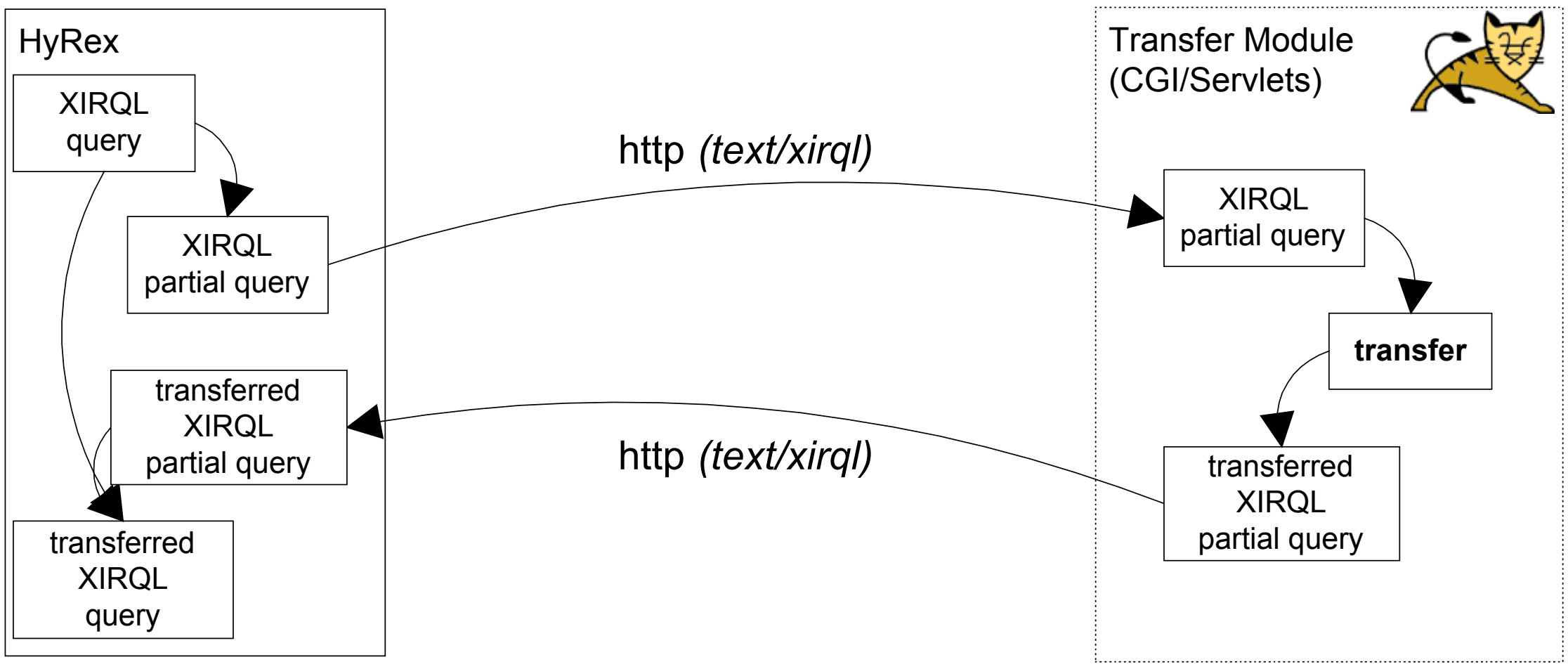

**Figure 18: Query Transfer**

The transfer service is implemented as a Java Servlet server (TomCat). This allows an easy integration of the class hierarchy for transfer modules (cf. section 7.6).

## 7.3 Transfer Use in ELVIRA

The result of the correlation process is a set of relations between pairs of terms where each such pair contains one term for each of the classification systems. Each of these relations is augmented with weight information that describes how close the pair of terms is related. In the project ELVIRA this is represented by a two-dimensional table where each column represents a term from the first classification system while each row represents a term from the second classification system. The cells of this table hold the weight information (cf. Table 2).

[13] http://ls6-www.informatik.uni-dortmund.de/~goevert/HyREX.html
[14] http://ls6-www.cs.uni-dortmund.de/ir/projects/carmen/wp7.html.de

| | Term A | Term B | Term C | Term D | Term E |
|---|---|---|---|---|---|
| Term 1 | 0 | 0 | 0.6 | 0 | 0.8 |
| Term 2 | 0 | 0 | 0.5 | 0.5 | 0 |
| Term 3 | 0.3 | 0 | 0.4 | 0 | 0.6 |
| Term 4 | 0.3 | 0.5 | 0 | 0 | 0.5 |
| Term 5 | 0 | 0.3 | 0 | 0.6 | 0 |
| Term 6 | 0.6 | 0 | 0.4 | 0 | 0 |

**Table 2: Resulting correlation table**

In order to transform a single term query from one classification system into another, the highest relevant terms can either be selected by applying an absolute threshold (e.g. all relations weighing 0.4 or more) or by selecting the maximum number of corresponding terms, using those with the highest values (the two terms with the highest weight). Alternatively, both of these methods can be combined. For example Term 3 could be transformed into [Term E], [Term E, Term C] or [Term E, Term C, Term A], depending on the threshold and the maximum number of desired result terms.

The weight information can, however, be employed more effectively for transformations of multi term queries where the query terms are not structured with Boolean operators, but rather "describe the topic of interest" in an associative manner. In this case, the model of a table can be used to sum up the row values of the query terms and afterwards apply the selection methods to the result row. This effectively makes the common aspects of the query term more relevant to the transformation process and allows for the identification of a new set of terms which in turn describe the same topic of interest without the problems of over expansion that are common to un-weighted transformation methods (cf. Table 3).

| | Term A | Term B | Term C | Term D | Term E |
|---|---|---|---|---|---|
| ⇒ Term 1 | 0 | 0 | 0.6 | 0 | 0.8 |
| ⇒ Term 2 | 0 | 0 | 0.5 | 0.5 | 0 |
| ⇒ Term 3 | 0.3 | 0 | 0.4 | 0 | 0.6 |
| Term 4 | 0.3 | 0.5 | 0 | 0 | 0.5 |
| Term 5 | 0 | 0.3 | 0 | 0.6 | 0 |
| Term 6 | 0.6 | 0 | 0.4 | 0 | 0 |
| Σ | 0.3 | | **1.5** | 0.5 | **1.4** |
| | | | ⇓ | | ⇓ |

**Table 3: Use of weight information in multi term transformations**

## 7.4 Transfers using Neural Nets

Similar to the multi-term-transformations based on co-occurrence (described in the previous section) the preferred use of neural nets is the transformation of unstructured groups of index terms. During the learning phase (see section 6.1) the network has memorized which term, or group of terms, from system *A* relates to which (group of) terms from system *B*. This relation is used as a transfer relation instead of an intellectual or statistical relation. So after the separation of the transformable query part the terms it consists of act as the input vector of the neural net. The resulting output vector forms the new, transformed query part, which will be combined with the rest of the users query as described in section 7.1.

Also the results of transformations with neural nets are different from those of intellectual or statistical transformations their integration into the retrieval process does not differ significantly. This allows a smooth integration of all three transformation models in one information retrieval system.

## 7.5 Transfer Use in ETB

The transfer modules will be tested with selected repositories of the pilot phase of the ETB network. Then they have to be applied to the specific conditions of the educational domain and of the types of documents being part of the repositories inside the network. Finally the most feasible modules will be chosen, and they need to be implemented into the final ETB software and tools package.

## 7.6 Class Hierarchy for Transfer Modules

To integrate the transfer relations into the retrieval in the sense of the 2-step approach described in sec. 1.1 we created a class hierarchy with Java classes and interfaces as a library for common use (cf. Figure 19). The generic transfer module is sub-classed by dividing different types of transfers, i.e. intellectual, statistical and neural transfers. The transfer rules can be stored in plain files, databases (RDBMS) or special sources like SIS-TMS (for intellectual cross-concordances). Some parameters allow users to influence the transfer process (e.g. relevance/weight, types of hierarchical relations).

Transfer requests are analysed and split up into atomic transfers that are handled by concrete classes querying the transfer rule sources. This allows processing complex queries in a very flexible way.

The programming of this library is in process, concrete implementations of requested transfer types follow on demand.

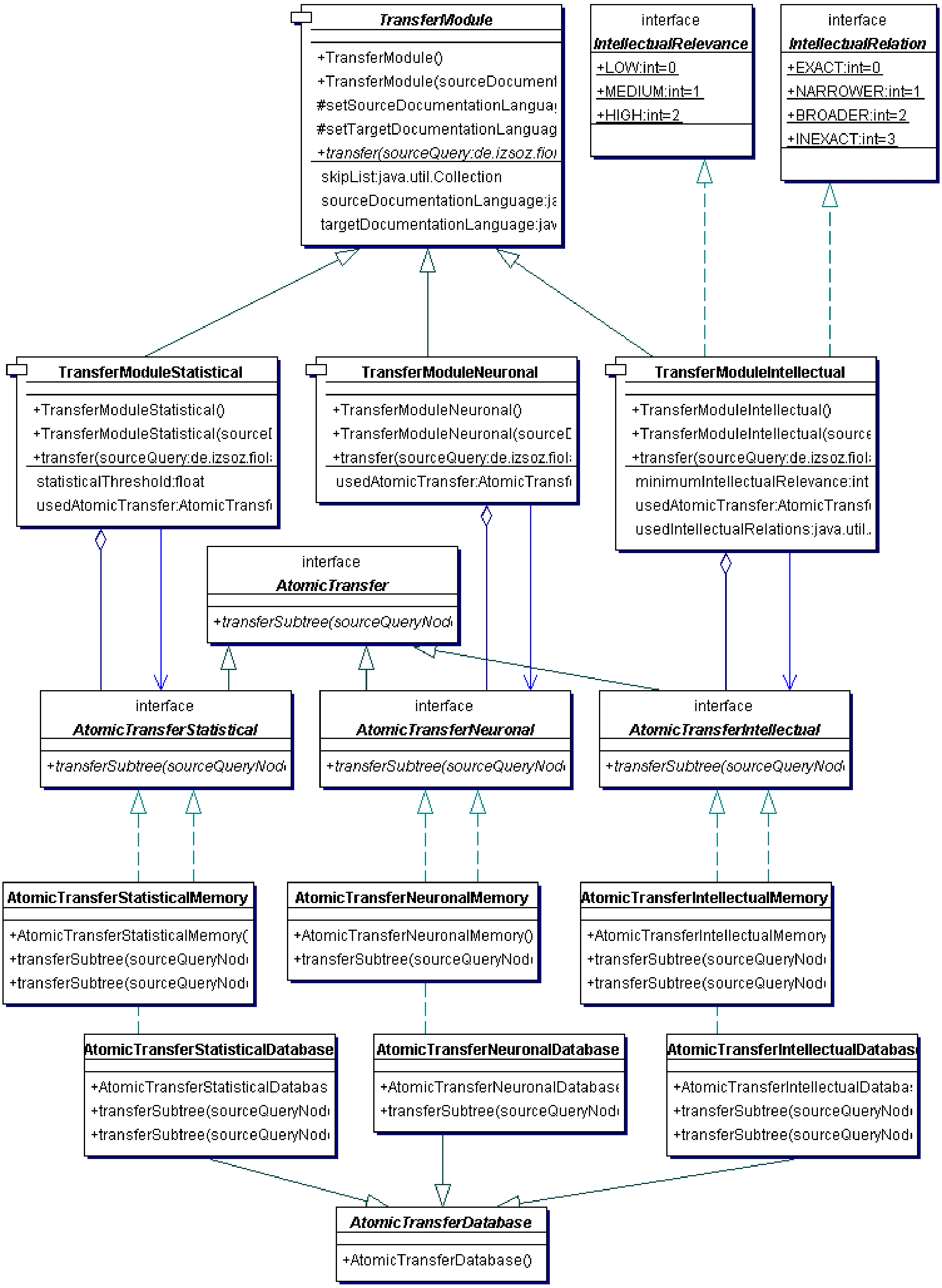

**Figure 19: Class Hierarchy for Transfer Module**

# 8 References

Bentz, Hans-Joachim; Hagström, Michael; Palm, Guenther (1998): Information Storage and Effective Data Retrieval in Sparse Matrices. In: Neural Networks 2 (4). S. 289-293.

Biebricher, P., Fuhr, N., Lustig, G., Schwantner, M. and Knorz, G. (1988): The Automatic Indexing System AIR/PHYS - From Research to Application. In 11th International Conference on Research & Development in Information Retrieval, (Grenoble, France, 1988), ACM Press.

Boughanem, M.; Soulé-Dupuy, C. (1998): Mercure at trec6. In: Voorhees/Harman 1998.

Burkart, Margarete (1997): Thesaurus. In: Buder, Marianne; Rehfeld, Werner; Seeger, Thomas; Strauch, Dietmar (ed.): Grundlagen der praktischen Information und Dokumentation. Ein Handbuch zur Einführung in die fachliche Dokumentationsarbeit. München, S. 160-179.

Chen, Hsinchun; Schuffels, Chris; Orwig, Richard (1996): Internet Categorization and Search: A Self-Organizing Approach. In: Journal of Visual Communication and Image Representation. 7 (1). S. 88-101.

Dorffner, Georg (1991). Konnektionismus. Von neuronalen Netzwerken zu einer natürlichen KI. Stuttgart.

Ferber, R., Automated Indexing with Thesaurus Descriptors (1997): A Co-occurrence Based Approach to Multilingual Retrieval. In: Research and Advanced Technology for Digital Libraries, (Pisa, 1997), Springer.

Grievel, L., Mutschke, P.; Polanco, X.: Thematic Mapping on Bibliographic Databases by Cluster Analysis (1998): A Description of the SDOC Environment with SOLIS. Knowledge Organisation, 22 (2). 8.

Haykin, S. (1994): Neural Networks: A Comprehensive Foundation.

Informationszentrum Sozialwissenschaften (ed.) (1999): Thesaurus for the Social Sciences / Thesaurus Sozialwissenschaften. German-English / Deutsch-Englisch. English-German / Englisch-Deutsch. (two volumes) Bonn.

Kluck, Michael (2001): Metadata and Handling of Heterogeneity as Central Means for the Development of an European School Portal. The Project European Schools Treasury Browser - ETB. Talk at the 7th Annual Meeting of the IuK Initiative Information and Communication of the Learned Societies in Germany »Cooperative Systems« Trier, Germany, March 11 - 14, 2001 (http://www.zpid.de/iuk2001/program/talks/Kluck-en.ppt; in German http://www.zpid.de/iuk2001/program/talks/Kluck-de.ppt)

Kohonen, Teuvo (1997): Self-Organizing Maps. Springer: Berlin et al.

Kohonen, Teuvo (1998): Self-organization of Very Large Document Collections: State of the art. In Niklasson, L.; Bodén, M.; Ziemke, T. (Hrsg.): Proceedings of ICANN ´98, 8th International Conference on Artificial Neural Networks, Springer: London. vol. 1, S. 65-74.

Krause, Jürgen (1996): Informationserschließung und -bereitstellung zwischen Deregulation, Kommerzialisierung und weltweiter Vernetzung; Schalenmodell. Bonn (IZ-Arbeitsbericht; Nr. 6).

Krause, Jürgen (2000): Virtual libraries, library content analysis, metadata and the remaining heterogeneity. In: 3rd International Conference of Asian Digital Library Conference (ICADL 2000), (Seoul 2000).

Krause, Jürgen; Marx, Jutta (2000): Vocabulary Switching and Automatic Metadata Extraction or How to Get Useful Information from a Digital Library. In: Information Seeking, Searching and Querying in Digital Libraries. Proceedings of the First DELOS Network of Excellence Workshop. Zurich, Switzerland, December 11-12, 2000. Zurich. pp. 133-134.

Mandl, Thomas (2000): Tolerant Information Retrieval with Backpropagation Networks. In: Neural Computing & Applications. Special Issue on Neural Computing in Human-Computer Interaction. 9 (4). Springer. S. 280-289.

Manecke, Hans-Jürgen (1997): Klassifikation. In: Buder, Marianne; Rehfeld, Werner; Seeger, Thomas; Strauch, Dietmar (Hg.): Grundlagen der praktischen Information und Dokumentation. Ein Handbuch zur Einführung in die fachliche Dokumentationsarbeit. München, S. 141-159.

Mitchell, Tom (1997) Machine Learning. Boston et al.

Mothe, Josiane (1994): Search Mechanisms Using a Neural Network Model. In: Intelligent Multimedia Information Retrieval Systems and Management. Proceedings of the RIAO ´94 (Rechenche d'Information assistée par Ordinateur). Rockfeller University. New York, 11.-13.10.94. S. 275-294.

Salton, Gerard (1987): Information Retrieval – Grundlegendes für Informationswissenschaftler. Hamburg – New York.

Strötgen, Robert; Kokkelink, Stefan (2001): Metadatenextraktion aus Internetquellen: Heterogenitätsbehandlung im Projekt CARMEN. In: Schmidt, Ralph (Hrsg.): Information Research & Content Management: Orientierung, Ordnung und Organisation im Wissensmarkt; 23. Online-Tagung der DGI und 53. Jahrestagung der Deutschen Gesellschaft für Informationswissenschaft und Informationspraxis e.V., DGI, Frankfurt am Main, 8. bis 10. Mai 2001; Proceedings. Frankfurt am Main: DGI 2001. (Tagungen der Deutschen Gesellschaft für Informationswissenschaft und Informationspraxis; 4), S. 56-66.

Zadeh, Lofti (1994): What is BISC? http://http.cs.berkeley.edu/projects/Bisc/bisc.memo.html#what_is_sc